\documentclass[12pt,a4paper,english,superscriptaddress,aps,nofootinbib]{revtex4}
\usepackage[utf8]{inputenc}
\usepackage[T1]{fontenc}
\usepackage{amsmath,amssymb,graphicx}
\makeatletter
\usepackage{babel}
\usepackage[active]{srcltx}
\usepackage{graphicx,color}
\usepackage{changebar}
\usepackage{hyperref}
\usepackage[T1]{fontenc}
\usepackage{esint}
\usepackage{multirow}
\usepackage{dsfont}
\usepackage{ae}
\usepackage{amsmath}
\usepackage{braket}
\usepackage{mathtools}
\usepackage{slashed}
\usepackage{empheq}
\usepackage{multirow}
\usepackage{graphicx}
\usepackage{amsfonts}
\usepackage{amsmath}
\usepackage{caption}
\begin{document}
\title{Shadow, absorption and Hawking radiation of a Schwarzschild black
	hole surrounded by a cloud of strings in Rastall gravity}

\author{Qian Li}
\affiliation{Faculty of Science, Kunming University of Science and Technology, Kunming, Yunnan 650500, China.}

\author{Chen Ma}
\affiliation{Faculty of Science, Kunming University of Science and Technology, Kunming, Yunnan 650500, China.}

\author{Yu Zhang}
\email{zhangyu\_128@126.com  (Corresponding author)}
\affiliation{Faculty of Science, Kunming University of Science and Technology, Kunming, Yunnan 650500, China.}

\author{Zhi-Wen Lin}
\affiliation{Faculty of Science, Kunming University of Science and Technology, Kunming, Yunnan 650500, China.}

\author{Peng-Fei Duan}
\affiliation{City College, Kunming University of Science and Technology, Kunming, Yunnan 650051, China.}



\begin{abstract}
This paper studies the black hole shadow, absorption cross section, and Hawking radiation of a massless scalar field in the background of a static spherically symmetric black hole spacetime that is surrounded by a cloud of strings in Rastall gravity. Specifically, the effects of the parameters $a$ and $\beta$ on the photon sphere and shadow radii are investigated. The results show that as the negative parameter $\beta$ decreases, the photon sphere and shadow radii change in an N-shape. In addition, the absorption cross section obtained after solving the massless Klein-Gordon equation is calculated using the sinc approximation and the partial waves method. We compare the absorption cross section obtained by the sinc approximation and the partial waves method,	and find it to be exceptionally consistent in the mid-to-high frequency region. Furthermore, the effects of parameters $a$ and $\beta$ on absorption are examined in detail. Finally, we study in detail the effects of the  parameters $a$, $\beta$ and $l$ on the Hawking radiation power emission spectrum of the considered black hole. It turns out that the string parameter $a$ always suppresses the power emission spectrum, indicating that such black holes live longer when the string parameter $a$ is increased while other parameters are fixed.
\end{abstract}

\maketitle


\section{Introduction}

	General relativity, proposed by Einstein in 1915 \cite{Einstein1914,Einstein1915}, is by far the most widely accepted theory of gravity. The predictions made therein have been tested and verified under weak or strong field conditions. Particularly, black holes, as one of the predictions, are arguably the most interesting and mysterious celestial bodies in our universe. The mystery of a black hole is that nothing, including light, can escape its event horizon. For the past few decades, the existence of black holes was only studied through indirect methods, until the first images of black holes appeared in 2019 \cite{Event2019}. This discovery provides many inspiring answers for our exploration of Einstein's  theory of general relativity and for testing other revised theories of gravity, taking our understanding of black hole physics a major step forward. However, the basic theory proposed by Einstein cannot explain some phenomena or solve certain fundamental problems, e.g., the singularity problem and the conjecture that the covariant divergence of the energy-momentum tensor may be non-zero.

To account for the special case where the covariant divergence of the energy-momentum tensor does not vanish,	Rastall \cite{Rastall1972} proposed a special modification of general relativity where the field equation is $ T^{\mu\nu}_{;\mu} = \lambda R^{,\nu}$ and $\lambda= 0 $ corresponds to the Einstein equation. An important feature of Rastall's gravity is that the field equation $ T^{\mu\nu}_{~~~;\mu} = \lambda R^{,\nu}$ is obtained directly by violating the normal conservation law, which does not rely  on the metric or palatini formalism  \cite{Gogoi2021}. Also, it is important to note that Rastall's gravity appears to be consistent with experimental observations in the context of cosmology \cite{Al-Rawaf1996}. Specifically, the observational data include, but are not limited to, the age of the universe, helium nucleosynthesis, and Hubble parameters. What is more interesting is that the modified gravity gives us a lot of novel and interesting results at the cosmological level. Besides, some attention has been focused on a debate, namely, whether Rastall gravity is equivalent to Einstein gravity.  	Visser  \cite{Visser:2017gpz} thought that the modified gravity proposed by Rastall  is a rearrangement of the matter sector of Einstein gravity. In other words, the geometrical part of the field equation is the  same in both theories, so we just need to  construct  a new energy-momentum tensor to fulfill the  ordinary conservation law. So the author claimed there is nothing new, such as gravity description, in Rastall proposal. Das et al.  \cite{Das:2018dzp} had a conclusion that in the framework of non-equilibrium thermodynamics (for homogeneous and isotropic FLRW black hole model), generalized Rastall gravity theory is equivalent to Einstein gravity theory.  However, other researchers disagree with Visser's ones, see for example the research of  Darabi and his colleagues \cite{Darabi:2017coc}  who support that  Rastall theory is not equivalent to Einstein gravity theory and give a simple example to prove that the claim proposed by Visser is incorrect. Moreover, they indicated that Rastall gravity theory is an "open" gravity theory in comparison to basic general relativity and has more compatible with observational cosmology. Hansraj et al. \cite{Hansraj:2018zwl} also discussed  this dispute and their results are  consistent with  Darabi et al. \cite{Darabi:2017coc}. In this work, they showed that Rastall gravity can satisfy the fundamental conditions for physically viable	model  whereas Einstein gravity doesn't fulfill its requirements (see \cite{Hansraj:2018zwl} for more detailed  discussion). Some works \cite{Ziaie:2019jfl,Moradpour:2017ycq,Li:2019jkv,Abbas:2018ffk}  have shown the difference between  Rastall gravity and  Einstein gravity from theoretical or cosmological perspectives. Finally, regardless of whether the Rastall gravity is equivalent to Einstein gravity,  Rastall gravity theory is worth studying or discussing because it faces a challenge from cosmological (astrophysical) observations.

String theory, on the other hand, holds that the fundamental unit of nature is not the point particle in particle physics, but an extended one-dimensional string. Letelier \cite{Letelier1979} proposed for the first time  that the source of the gravitational field could be a cloud of strings, and gave an exact solution for a Schwarzschild black hole surrounded by a collection of strings in the context of Einstein's general relativity. In addition, black holes that treat a cloud of strings as the source of the gravitational field in the modified gravity have been studied \cite{Herscovich2010,Toledo2019,Morais2018,Li2021,Chen:2018szr}. For instance, Cai and Miao proposed a black hole solution in which a cloud of strings is the source of the gravitational field of a Schwarzschild black hole in the context of Rastall gravity \cite{Cai2020}. The authors also analyzed fundamental thermal properties, quasinormal modes of gravitational perturbations, area spectra \cite{Setare:2003bd,Setare:2004uu}, and entropy spectra.

The experimental results reported by the Event Horizon Telescope Collaboration \cite{Event2019} not only directly prove the existence of black holes, but also allow us to directly observe the shadows of black holes. The theoretical analysis of black hole shadows has a long history. For example, Synge \cite{Synge1966} discussed the shadows of Schwarzschild spacetime, and  Bardeen et al. \cite{Bardeen} analyzed the shadows of Kerr black holes. In addition to shadow analysis performed in basic general relativity, it extends to various modified forms of gravity or arbitrary-dimensional spacetime. Abbas and Sabiullah \cite{Abbas:2014oua} studied  the structure of timelike  as well as null geodesics of  regular Hayward black hole and found the massive particles, which move along timelike geodesics path, are dragged toward the black hole.	To the best of our knowledge, numerous studies \cite{Amarilla2010,Sharif2016,Amir2019,Babar2020,Konoplya20200,Konoplya20201,Anacleto2021,Cai2021,Zhang2021,Long:2019nox,Long:2020wqj} have  been devoted to studying the shadows of black holes under various modified gravity. More concretely, Gyulchev et al. \cite{Gyulchev2019,Gyulchev2018,Nedkova2013} analyzed the shadows cast by different rotating traversable wormholes. Interestingly, the near horizon geometry is determined by the shadow cast by the black hole. Instead, the trajectory of light is affected by the plasma surrounding the black hole. This causes the geometric size and shape of the shadow  on  the Kerr spacetime  to change  \cite{Perlick2017}. In general, gravitational light deflection causes black hole shadows, and the trajectory of a photon in a vacuum depends on its impact parameter \cite{Javed2021}. Therefore, we cannot ignore the role of the impact parameter in shadow formation.

Due to the special properties of black holes, we cannot directly study their internal structure. However, black hole is not an isolated system because it interacts with its surrounding environment, such as absorption, scattering and Hawking radiation. These interactions can convey information about the interior of the event horizon. In particular, as one of the interactions, the absorption cross section of black holes has received extensive attention from researchers. That's because one of the most useful and efficient ways to understand the properties of a black hole is to analyze the absorption of matter waves and the test field around the black hole. This series of studies began in the 1970s \cite{Matzner1968,Mashhoon1973,Starobinski1974,Fabbri1975,Ford1975,Page1976,Sanchez19781}. During that period, Sanchez found that the absorption cross section of Schwarzschild spacetime for scalar waves oscillates around the geometric capture cross section. About twenty years later, Das et al. \cite{Das1997} presented a key result that, in the low-energy regime, the absorption cross section of a coupled massless scalar field is equal to its event horizon area. Consequently, the literature on this particular topic has proliferated over the past few decades, covering various fields of research and several revision theories \cite{Higuchi2001,Kanti2002,Jung2004,Grain2005,Crispino2007,Crispino2009,Macedo2014,Huang2015,2018,Huang2019,Anacleto2020,Magalhaes2020,Junior2020,Benone2018,Li2022}.

Furthermore, Hawking predicted that black holes are thermal systems, like black bodies, and then have associated temperature and entropy. Based on the analysis of quantum field dynamics in the context of curved space-time, Hawking pointed out that black holes emit radiation, known as Hawking radiation, from their event horizons \cite{Hawking1975,Hawking1976}. Intriguingly, Hawking radiation depends on the type of particle and the geometry of the black hole. This is because the Hawking temperature $T_{BH}=\frac{f'(r_{+})}{4\pi}$ is one of the influencing factors. Moreover, Yale \cite{Yale2011} has analyzed the Hawking radiation of particle scalars, fermions and bosons spin-1 using the tunneling method. In recent years, a large body of literature \cite{Chen:2008ra,Konoplya2020,Harmark2010,Kanti2015,Pappas2016,Miao2017,Javed2019,2020,Guo2020,Ali2021,Slavov:2012mv} has emerged on Hawking radiation on various modified gravity, including high-dimensional black holes.

This paper investigates the black hole shadow, absorption cross section and Hawking radiation of the test scalar field of a Schwarzschild black hole surrounded by a cloud of strings in Rastall gravity. Specifically, Cai and Miao \cite{Cai2020} presented the corresponding quasinormal modes of odd parity gravitational field by the  WKB approximation.  On this basis, our research contributes to further understanding of this black hole and its physical characteristics.

This paper is organized as follows. The second section outlines the basic information of the black hole solution, that is, a Schwarzschild black hole surrounded by a cloud of strings in the context of Rastall gravity, and also gives the meaning of the influencing parameters. The third part is devoted to the derivation of massless scalar equations and the analysis of related effective potentials. Section 4 analyzes the radius of the photon sphere and the shadow radius of the black hole in detail. Next, the absorption cross section of the scalar field is calculated using the sinc approximation and the partial wave method, and the effects of the parameters are also investigated. Section 6 gives the expression of Hawking radiation and the corresponding results for the Hawking radiation power emission spectra. The last section contains the summary and conclusions. Besides, we use the natural unit that $c = G = \hbar = 1$ in this paper.

	\section{The solution of a Schwarzschild black hole surrounded by a cloud of strings in Rastall gravity}

The field equations of the Rastall gravity \cite{Rastall1972} are as follows,

\begin{equation}\label{QQ}
G_{\mu\nu} + \beta g_{\mu\nu} R= \kappa T_{\mu\nu} ,
\end{equation}
\begin{equation}\label{QW}
T^{\mu\nu}_{~~~;\mu} = \lambda R^{,\nu} ,
\end{equation}
where $\kappa$ and $\lambda $ represent the Rastall gravitational coupling constant and the Rastall parameter, respectively. Moreover, $\beta$ is defined as the product of these two parameters, i.e., $\beta \equiv \kappa \lambda $. From the above equations we have that
\begin{equation}\label{Q3}
R= \frac{\kappa}{4\beta-1}T ,
\end{equation}
\begin{equation}\label{4}
T^{\mu\nu}_{~~~;\mu} = \frac{\kappa}{4\beta-1}T^{,\nu} ,
\end{equation}
where $R$, $T$ denote the Ricci scalar and the trace of the energy-momentum tensor, respectively. Besides, $\kappa=\frac{4\beta-1}{6\beta-1} 8 \pi$ under the Newtonian limit \cite{Moradpour2016}. It can be seen from the above equations that, the Einstein gravity is recovered and the energy-momentum tensor is conserved when the Rastall gravity parameter $\lambda$ vanishes, i.e., $\beta=0$.

We consider the case where the metric is static and spherically symmetric,
\begin{equation}\label{Q5}
\text{d}s^2=-f(r)\text{d}t^2+f^{-1}(r)\text{d}r^2+r^2\text{d}\theta^2+r^2 \sin^{2}\theta\text{d}\phi^2
\end{equation}
with the metric \cite{Cai2020}
\begin{equation}\label{Q6}
f(r)=1-\frac{2M}{r}+\frac{4a(\beta-\frac{1}{2})^{2}}{(8\beta^{2}+2\beta-1)r^{\frac{4\beta}{2\beta-1}}}.
\end{equation}

It is worth noting that the Rastall theory should satisfy the Newtonian limit \cite{Moradpour2016}. Therefore, the cases $\beta=\frac{1}{6}$ and   $\beta=\frac{1}{4}$ are not allowed. The parameter $a$ needs to satisfy a specific constraint, namely $a\equiv\kappa b$ where $b$ is a constant of integration associated with a cloud of strings. Specifically, $\beta $ and $a$ represent the influence of the Rastall gravity and the string, respectively. Consequently, the Rastall gravity is converted to Einstein gravity when $\beta=0$. Meanwhile, when $a$ equals to 0, the Schwarzschild spacetime is restored.

\section{scalar wave equation }

The massless scalar field  $\Psi$  governed by the massless Klein-Gordon equation in curved spacetime can be formulated as
\begin{equation}\label{Q11}
\frac{1}{\sqrt{-g}}\partial_{\mu}(\sqrt{-g}g^{\mu\nu}\partial_{\nu})\Psi=0,
\end{equation}
and then the massless scalar field  $\Psi$ can be decomposed as follows
\begin{equation}\label{Q8}
\Psi_{\omega l m}=\frac{\psi_{\omega l(r)}}{r} P_{l}(\cos\theta)e^{ -i \omega t},
\end{equation}
where $P_{l}(\cos\theta)$ denotes the Legendre polynomial, $l$ and $m$ represent the corresponding  angular quantum number and  magnetic quantum number, respectively. In addition, the function $\Psi_{\omega l}$ satisfies the following ordinary differential equation,
\begin{equation}\label{Q9}
f(r)\frac{d}{dr}[f(r)\frac{d\psi_{\omega l}}{dr}]+[\omega^{2}-V_{eff}(r)]\psi_{\omega l}=0,
\end{equation}
where $V_{eff}(r)$  stands for the corresponding effective potential that is defined as
\begin{equation}\label{Q10}
V_{eff}(r)=f(r)(\frac{1}{r}\frac{df(r)}{dr}+\frac{l(l+1)}{r^2}).
\end{equation}

Moreover, by substituting the metric in the effective potential, the specific potential is reformulated as
\begin{equation}\label{Q101}
\begin{aligned}
& V_{eff}(r)=\Bigg(1-\frac{2M}{r}+\frac{4a(\beta-\frac{1}{2})^{2}}{(8\beta^{2}+2\beta-1)r^{\frac{4\beta}{2\beta-1}}}\Bigg)\times\\
&
\Bigg(\frac{l(l+1)}{r^{2}}+\frac{2M}{r^{3}}-\frac{16 a(-\frac{1}{2}+\beta^{2} )r^{-2-\frac{4\beta}{-1+2\beta}}}{(-1+2\beta)(-1+2\beta+8\beta^{2})} \Bigg).
\end{aligned}
\end{equation}

Additionally, we define the following tortoise coordinate change
\begin{eqnarray}
r_{*}=\int \frac{dr}{f}.
\end{eqnarray}

Consequently, the equation (\ref{Q9}) is equivalent to
\begin{eqnarray}\label{Q12}
\frac{d^2{\psi}}{dr_{*}^{2}}+(\omega^{2}-V_{eff})\psi=0.
\end{eqnarray}

Note that both the metric $f(r)$ and the effective potential $V_{eff}(r)$ are divergent when $\beta$ is set to -0.5. Besides, to satisfy the condition that the effective potential $V(r) \rightarrow 0$ when $r\rightarrow \infty$, we have $\beta < \frac{1}{6}$. Hence, the domain of $\beta$ should be in $(-0.5, \frac{1}{6})$. Moreover, due to the condition $ a \equiv \kappa b$, the domain of $a$ strictly relies on the positivity (negativity) of the parameter $\beta$. For $\beta < 0$, the barrier of the effective potential $V_{eff}(r)$ disappears as the parameter $a$ approaches to 1. Accordingly, the domain of the parameter $a$ is set to $[0,1)$. In contrast, for $\beta > 0 $, when the parameter $a$ is larger, a black hole surrounded by a cloud of strings in Rastall gravity has no event horizon. For instance, when $\beta=0.1$, the domain of $a$ is set to $[0,0.3]$.

Fig.\ref{FIG1} shows the  behaviour of the effective potential $V_{eff}(r)$ with respect to $r$ for different angular quantum numbers $l$ when $a=0.1$, $\beta=\frac{1}{10}$. We find that the peak value of the effective potential increases when the angular quantum number $l$ is increased. Furthermore, the potential $V_{eff}(r)$ first increases, then decreases, and finally tends to  zero at $r\rightarrow\infty$.

As shown in Fig.\ref{FIG2}, to compare the effects of parameters $a$ and $\beta$ on the effective potential $V_{eff}(r)$, we depict the behaviour of $V_{eff}(r)$ with respect to $a$ and $\beta$ when $\beta < 0$ and $ \beta > 0$, respectively. Specifically, for $\beta > 0$, i.e., when the parameter $\beta$ is fixed to $\frac{1}{10}$,  the barrier height of the effective potential decreases as the string parameter $a$ increases. It is clear that the peak of the effective potential becomes smaller and shifts to the right side as $a$ increases. Next, we vary the Rastall parameter $\beta$ and fix $a$ to $0.1$. It can be seen that with the increase of $\beta$, the peak value of the effective potential decreases, and the position of the peak value does not change much compared with the case where the string parameter $a$ changes.

Meanwhile, when $\beta < 0 $, one can see that for the same value of $a$, the barrier height of the potential first increases and then decreases with decreasing $\beta$. Also, the peak position firstly shifts to the left and then to the right. Furthermore, when the parameter $a$ is varied, at the same value of the Rastall parameter $\beta$, the barrier height decreases and the peak position shifts to the right as the parameter $a$ increases.

\begin{figure}\label{FIG1}
	\centering
	  \includegraphics[height=5cm]{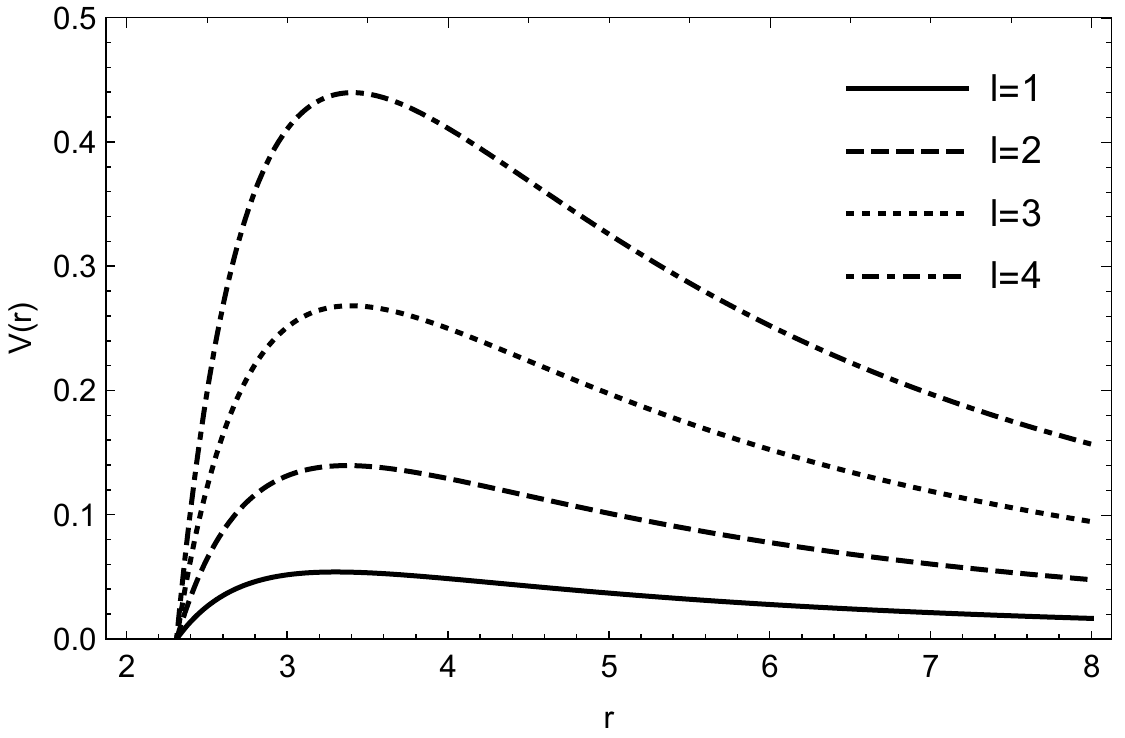}\quad
	\caption{\label{FIG1}  The variation of the effective potential with $r$ for $l=1,2,3,4$, with the fixed $ a=0.1$, $\beta=\frac{1}{10}$ and $M=1$.
}\end{figure}
\begin{figure*}\label{FIG2}
	\centering
	  \includegraphics[height=5cm]{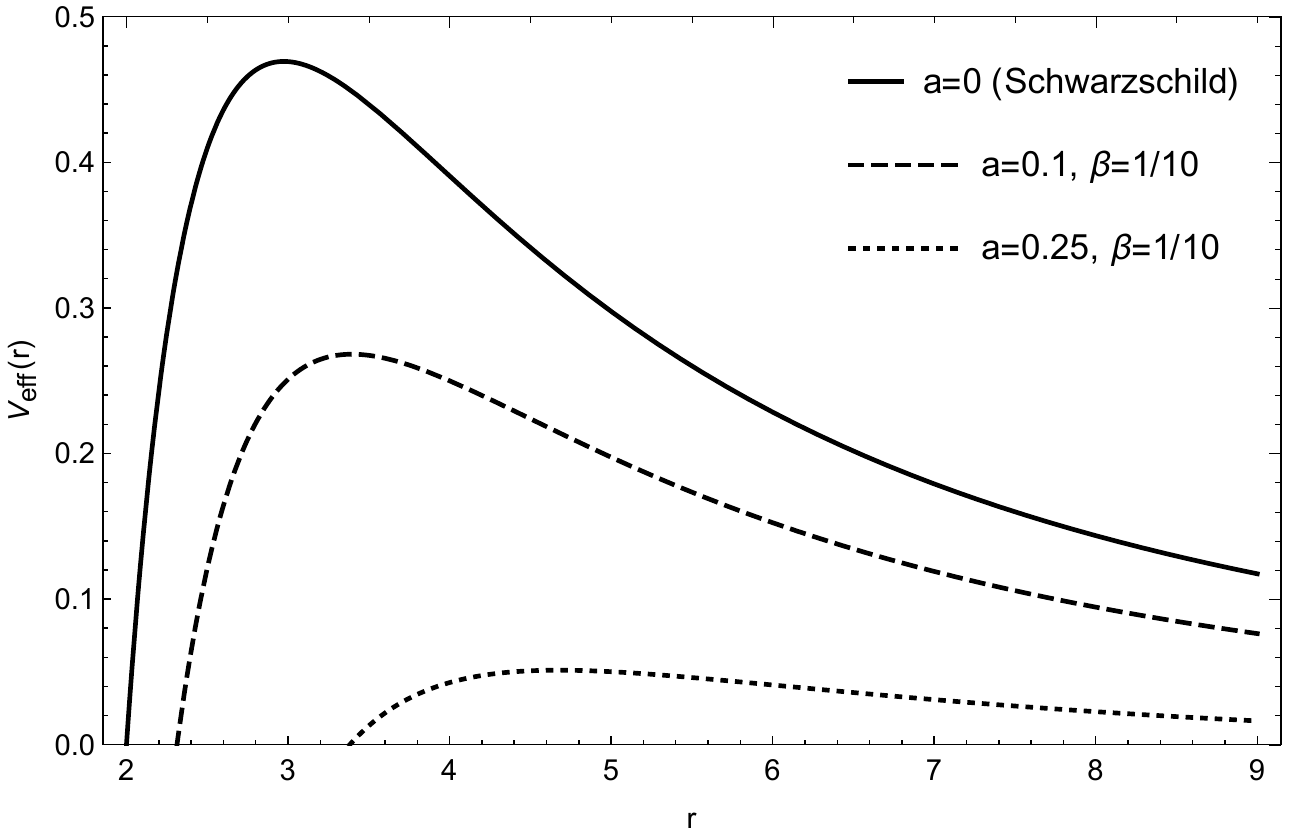}\quad
	  \includegraphics[height=5cm]{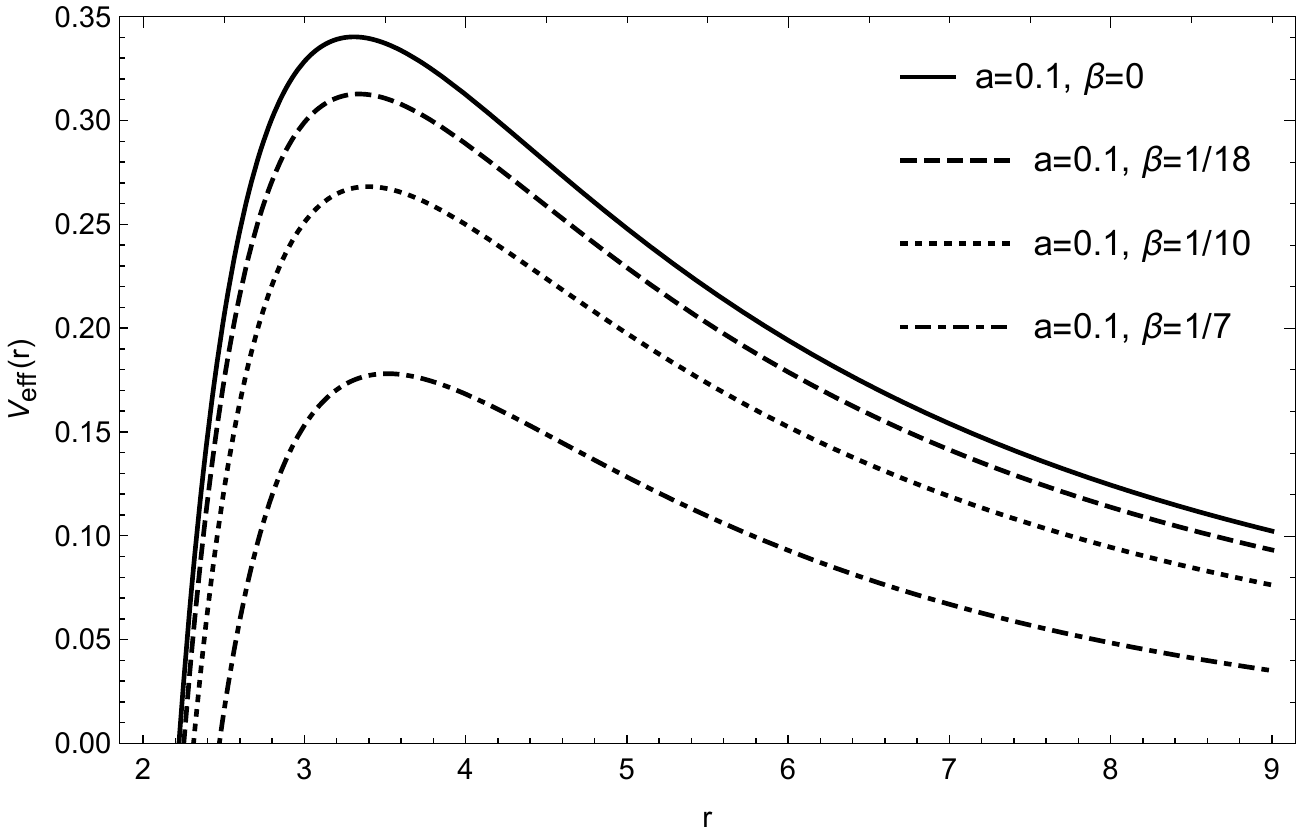}\quad
	  \includegraphics[height=5cm]{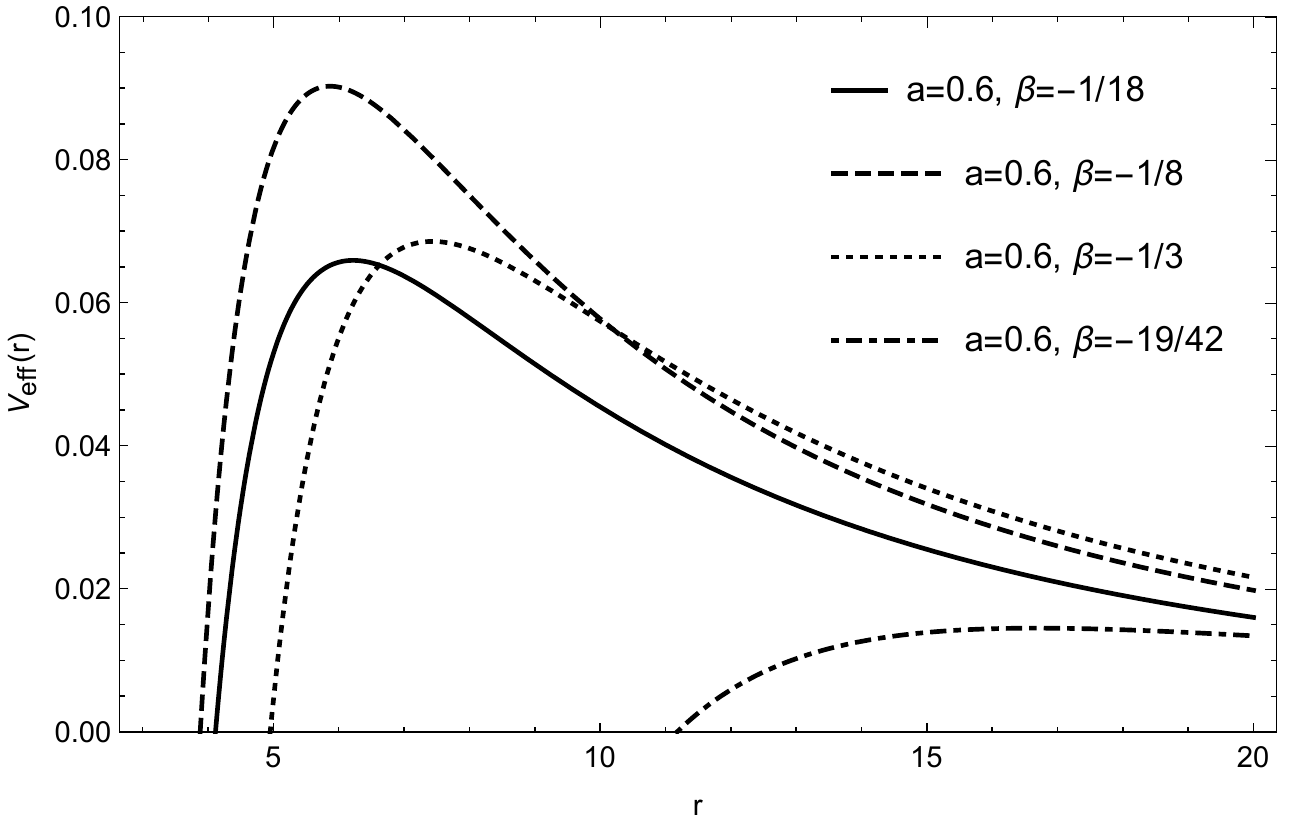}\quad
	  \includegraphics[height=5cm]{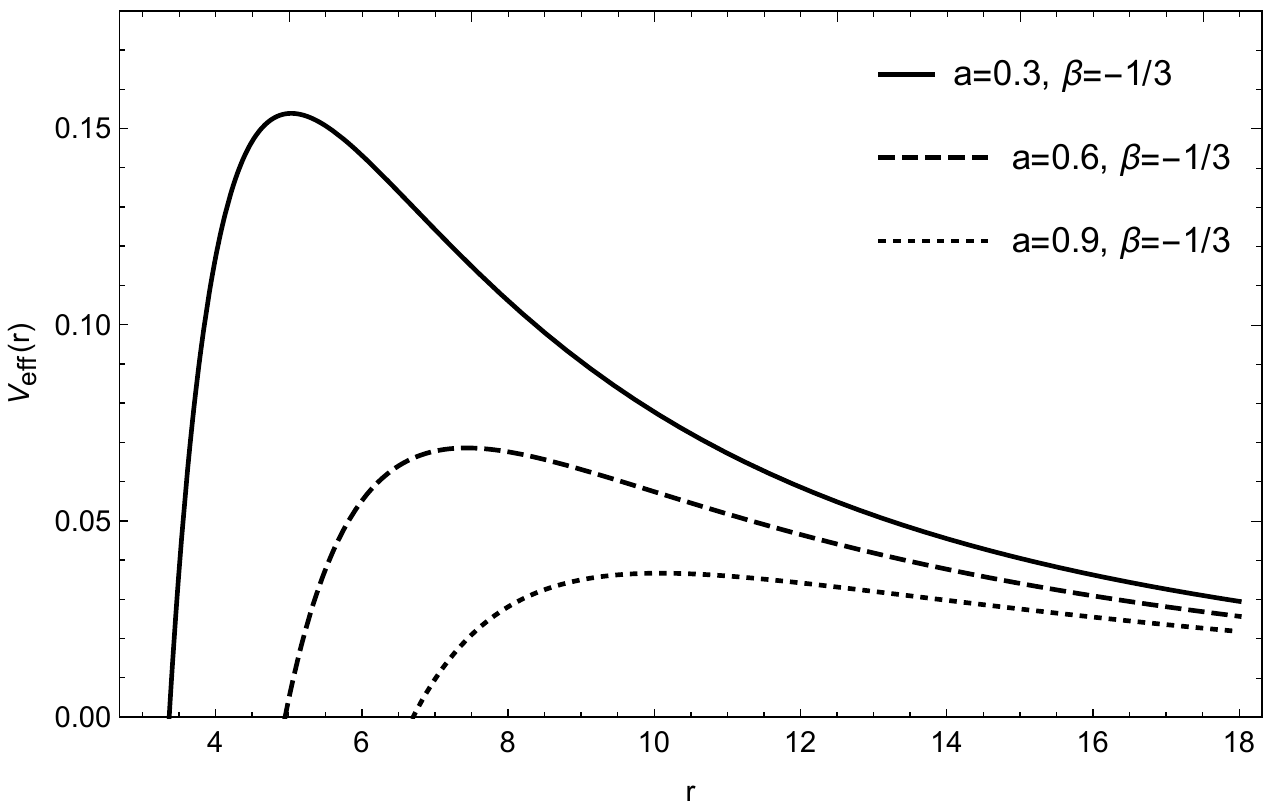}\quad
	\caption{\label{FIG2} The effective potential for the different variables with $M=1$, $l=3$. For $\beta > 0$, the picture in the upper left corner is that the parameter $a$ is variable when $\beta =\frac{1}{10}$. The plot in the upper right corner is that the parameter $\beta$ is variable when $a =0.1$. For $\beta < 0$, the picture in the lower left corner is that the parameter $\beta $ is variable with $a=0.6$. the figure in the lower right corner is that the parameter $ a $ is variable for fixed $\beta=-\frac{1}{3}$.
	}
\end{figure*}

We add some boundary conditions for the Sch$\ddot{\text{o}}$rding-like equation ($\ref{Q12}$) because we are interested in the absorption cross section and Hawking radiation. Near the horizon regime and at infinity, one can find that $\psi_{\omega l}(r_{*})$ need to satisfy the following boundary conditions

\begin{eqnarray}\label{E_nled3}
\psi_{\omega l}(r_{*})\sim\left\{
\begin{aligned}
I_{\omega l} e^{-i\omega r_{*}}+R_{\omega l} e^{i\omega
	r_{*}},~
r_{*}~\rightarrow+\infty, \\
T_{\omega l} e^{-i\omega r_{*}},
~r_{*}\rightarrow-\infty,
\end{aligned}
\right.
\end{eqnarray}
where $R_{\omega l}$ and $T_{\omega l}$  in the action denote reflection and transmission coefficients, respectively. Due to the conservation of flux,
$R_{\omega l}$ and $T_{\omega l}$ satisfy the following constraint
\begin{eqnarray}\label{E_nled4}
\left|R_{\omega l}\right|^2+\left|T_{\omega l}\right|^2=\left|I_{\omega l}\right|^2.
\end{eqnarray}

Furthermore, the phase shift $ \delta_{l} $ can be defined as
\begin{eqnarray}\label{E_nled5}
e^{2 i\delta_{l}}=(-1)^{l+1} R_{\omega l}/ I_{\omega l}.
\end{eqnarray}

Next, we will discuss the black hole shadows, absorption cross section and Hawking radiation based on the last two sections.

	\section{shadows}

In this section, we investigate the role of the Rastall parameter $\beta$ and the string parameter $a$ on the shadow radius of a black hole enclosed by a cloud of strings in Rastall gravity. Moreover, the results will be compared to those of Schwarzschild spacetime (i.e. $a=0$) and Einstein gravity (i.e. $\beta=0$), respectively.

The photon trajectories of a black hole surrounded by a cloud of strings in Rastall gravity can be represented by null geodesics \cite{Abbas:2014oua,Azam:2017adt}. The Lagrangian of  geodesic equations for the curve spacetime have the following form
\begin{eqnarray}\label{Q13}
0=-f(r)\dot{t}^{2}+\frac{1}{f(r)}\dot{r}^{2}+r^{2}\dot{\theta}^{2}+r^{2}\sin^{2}\theta\dot{\phi}^{2},
\end{eqnarray}
where the overdot symbol  denotes the differentiation with respect to the affine parameter $\tau$.	Without loss of generality, we consider an analysis restricted to the equatorial plane, i.e., $\theta=\frac{\pi}{2}$. By using the Euler-Lagrange equation, the $t$ and $\phi$ coordinates are expressed as,
\begin{eqnarray}\label{Q14}
\dot{t}=\frac{E}{f(r)},
\end{eqnarray}
\begin{eqnarray}\label{Q15}
\dot{\phi}=\frac{L}{r^2},
\end{eqnarray}
where $E$, $L$ are motion constants, representing the energy and angular momentum of the massless test particle, respectively.

Hence, by substituting Eq. (\ref{Q14}) and  Eq. (\ref{Q15}) in the Lagrangian equation (\ref{Q13}), the Lagrangian expression can be written as
\begin{eqnarray}\label{Q16}
\dot{r}+f(r)\big(\frac{L^2}{r^2})=E^{2},
\end{eqnarray}
furthermore, we define
\begin{eqnarray}\label{Q17}
V=f(r)\frac{L^2}{r^2},
\end{eqnarray}
where $V$ stands for the effective potential of the massless test particle. Besides, the null-like geodesics of the equatorial circular motion in static spherically symmetric spacetime should satisfy the conditions $\dot{r}=0$ and $\ddot{r}=0$. Consequently, we have $V=E^2$ and $\frac{dV}{dr}=0$, indicating the stability of circular null geodesics. The equations $V(r_p)=0$ and $V^{'}(r)_{|r=r_p}=0$ \cite{Setare:2010zd} represent the circular orbit of the photon, that is, the photon sphere radius $r_{ p}$.

Moreover, the critical impact parameter $b_{c}$ can be expressed as
\begin{eqnarray}\label{Q18}
b_{c}=\frac{L}{E}=\frac{r_{p}}{\sqrt{f(r_{p})}},
\end{eqnarray}
\begin{eqnarray}\label{Q19}
f^{'}(r_{p})(r_{p}-2f(r_{p}))=0.
\end{eqnarray}
On the other hand, the black hole shadow radius $r_{s}$ is represented by the celestial coordinates $(x,y)$ as follows
\begin{eqnarray}\label{Q20}
r_{s}=\sqrt{x^2+y^2}=\frac{r_{p}}{\sqrt{f(r_{p})}}.
\end{eqnarray}

Specifically, the effects of the parameters $a$ and $\beta$ on the photon sphere and shadow radii are shown in Table \ref{R2}. For $\beta > 0$, from Table \ref{R2} one can see that for  fixed $a=0.1$ ($\beta=\frac{1}{10}$), the photon sphere and shadow radii increase as the parameter $\beta$ ($a$) increases. Furthermore, as the string parameter $a$ tends to 0.3 when $\beta=\frac{1}{10}$, the black hole shadow radius increases rapidly. For $\beta < 0$, when we set $a=0.3$, we observe that the photon sphere and shadow radii first increase, then decrease and finally increases as the parameter $\beta$ decreases. A possible reason is that the metric $f(r)$ is not a monotonic function of the Rastall parameter $\beta$ in the range $-0.5 < \beta < 0$. Therefore, when the parameter $\beta$ is set to $-\frac{1}{3}$, the photon sphere and shadow radii increase as $a$ approaches to its parameter maximum.

\begin{table*}
	\centering
	\caption{The  photon sphere  and shadow radii with $M=1$.}%
	\label{R2}%
	\addtolength\tabcolsep{10pt}
	\begin{tabular}
		[c]%
		{|p{0.6in}|p{0.6in}|p{0.6in}|p{0.6in}|p{0.6in}|p{0.6in}|}%
		\hline
		\multicolumn{3}{|c|}{$a=0.1$} & \multicolumn{3}{|c|}{$\beta=\frac{1}{10}$}\\\hline
		$\beta $ & $r_{p}$ & $r_{s}$ & $a$ & $r_{p}$ & $r_{s}$ \\\hline
		0 & 3.333 & 6.086 & 0 & 3.000 & 5.190  \\
		3/100 & 3.346 & 6.195 & 0.1 & 3.422 & 6.829  \\
		3/50 & 3.367 & 6.371 & 0.15 & 3.716 & 8.212  \\
		9/100 & 3.405 & 6.676 & 0.25 & 4.696 & 15.437  \\
		3/25 & 3.466 & 7.281 & 0.3 & 5.777 & 50.951  \\
		\hline
		\multicolumn{3}{|c|}{$a=0.3$} & \multicolumn{3}{|c|}{$\beta=-\frac{1}{3}$}\\\hline
		$\beta $ & $r_{p}$ & $r_{s}$ & $a$ & $r_{p}$ & $r_{s}$ \\\hline
		
		-1/18 & 4.177 & 8.113 & 0.05 & 3.318 & 5.786  \\
		-1/8 & 4.179 & 7.810 & 0.15 & 3.989 & 7.035  \\
		-1/3 & 5.076 & 9.060 & 0.35 & 5.457 & 9.772 \\
		-2/5 & 6.313 & 11.162 & 0.65 & 7.916 & 14.362 \\
		-19/42 & 9.723 & 17.041 & 0.95 & 10.619 & 19.415\\
		\hline
	\end{tabular}
\end{table*}

\section{absorption cross section}

In this section, we calculate the absorption cross section using two methods, viz., the sinc approximation method and the partial waves method where the gray-body factor is calculated by sixth-order WKB method. Besides, we view the capture cross section as a reference. It is known that the absorption cross section at the low-frequency and high-frequency limits can be calculated by different analytical approximations. The total absorption cross section of massless scalar waves in an arbitrary-dimensional general spherically symmetric black hole inclines to its area \cite{Das1997} in the low-frequency regime, which is the event horizon of the black hole. In the high-frequency regime, the total absorption cross section of the massless scalar field converges to the geometric capture cross section, described by the following null geodesics
\begin{eqnarray}\label{Q21}
\sigma_{geo} \equiv \pi b^{2}_{c},
\end{eqnarray}
where $b_{c}$ denotes the above critical impact parameter.

\subsection{sinc approximation}
Sanchez \cite{Sanchez19781} proposed that in the high-frequency regime, the total absorption cross-section oscillates near the above-mentioned capture cross section  $(27/4)\pi r^{2}_{s}$, where $r^{2}_{s}=2M$, and has an interval of oscillation peaks, $\Delta=\frac{2}{\sqrt{27}M}$. In addition, Sanchez also presented the following analytical approximation of the absorption cross section
\begin{eqnarray}\label{Q22}
\sigma_{San}=\frac{27\pi}{4}-\frac{A}{\omega r_{s}}\sin\pi{(3\sqrt{3})}(\omega r_{s}+B),
\end{eqnarray}
which has the best fit when  $A= 1.14 \sim \sqrt{2}$ and $B<10^{-4}$.

Furthermore, the Sanchez approximation was generalized by D$\acute{e}$canini et al. to static spherically symmetric spacetimes of arbitrary dimensions. D$\acute{e}$canini et al. \cite{Yves2011} showed that in the eikonal state, the fluctuation of the absorption cross section was completely and very simply described by the properties of the null unstable geodesics located on the photon sphere. Important characteristics are the orbital period and the Lyapunov exponent. Specifically, the sinc approximation of the absorption cross section in a d-dimensional static and spherically symmetric black hole is given by
\begin{eqnarray}\label{Q23}
\sigma \approx  \sigma_{geo} +\sigma_{abs}^{osc},
\end{eqnarray}
where the oscillation part of the absorption, i.e., $\sigma_{abs}^{osc}$, is expressed as
\begin{eqnarray}\label{Q24}
\sigma_{abs}^{osc} \equiv  (-1)^{d-3}4(d-2)\pi \eta_{c} e^{-\pi\eta_{c}} \operatorname{sinc}({\frac{2\pi r_{c} \omega}{\sqrt{f(r_{c})}}) \sigma_{geo}},
\end{eqnarray}
with $\operatorname{sinc}(x)$ denoting the sine cardinal
\begin{eqnarray}\label{Q25}
\operatorname{sinc}(x) \equiv \frac{\sin x}{x},
\end{eqnarray}
and $d$ representing the dimension of the black hole. Besides $2\pi\frac{r_{c}}{\sqrt{f(r_{c})}}
= 2\pi b_{c} $ indicates the orbital period of the black hole on the photon sphere \cite{Decanini2010}. The parameter $\eta_{c} $ for measuring the instability of the circular orbit on the photon sphere is defined as
\begin{eqnarray}\label{Q26}
\eta_{c} =\frac{1}{2}\sqrt{4f(r_{c})-2r^{2}_{c} f^{''}(r_{c})},
\end{eqnarray}
for instance, the sinc approximation of the absorption cross section of a Schwarzschild black hole at the high-frequency limit is written as
\begin{eqnarray}\label{Q27}
\sigma \approx  \sigma_{geo} - 8\pi  e^{-\pi} \operatorname{sinc}[{2\pi (3\sqrt{3}M)\omega}] \sigma_{geo}.
\end{eqnarray}

\subsection{Partial wave approach }
We consider that the field $\Phi$,  which is purely ingoing waves at the event horizon, is the sum of the monochromatic incident plane wave $\Phi^{I}$ and outgoing scattered wave $\Phi^{S}$ in the far-field, that is,
\begin{eqnarray}\label{Q28}
\Phi\sim \Phi^{I}+\Phi^{S}.
\end{eqnarray}

Without loss of generality, we assume  that the direction of wave propagation is along the $z$-axis. Accordingly, the monochromatic incident plane wave $\Phi^{I}$ and the outgoing scattered wave $\Phi^{S}$ are respectively defined as
\begin{eqnarray}\label{Q29}
\Phi^{I}= e^{-i\omega(t-z)},
\end{eqnarray}
\begin{eqnarray}\label{Q30}
\Phi^{S}= \frac{1}{r} \hat{f}(\theta) e^{-i\omega(t-r)},
\end{eqnarray}
where $\hat{f}(\theta)$ denotes the scattering amplitude. Moreover, $e^{i\omega z}$ can be decomposed as \cite{Futterman1988}
\begin{eqnarray}\label{Q31}
e^{i\omega  z}= \sum_{l=0}^{\infty}(2l+1)i^{l}j_{l}(\omega r)P_{l}(\cos\theta),
\end{eqnarray}
with $j_{l}(.)$ representing the spherical Bessel function.

Hence, Eq. (\ref{Q29}) in the far-field can be rewritten as follows,
\begin{eqnarray}\label{Q32}
\Phi^{I}\sim \frac{e^{-i\omega t}}{r} \sum_{l=0}^{\infty} C_{\omega l}\big(e^{-i\omega r}+ e^{-i \pi(l+1) e^{i\omega r}}\big)P_{l}(\cos\theta),
\end{eqnarray}
where $C_{\omega l}$ is given by
\begin{eqnarray}\label{Q33}
C_{\omega l}=\frac{(2l+1)}{2i\omega}e^{i \pi(l+1)}.
\end{eqnarray}

The field solution $\Phi$ depends on the boundary conditions (\ref{E_nled3}). This means that the ingoing part of $\Phi$ should match the incident plane wave $\Phi^{I}$. Therefore we obtain
\begin{eqnarray}\label{Q34}
\Phi=\frac{e^{-i\omega t}}{r} \sum_{l=0}^{\infty} C_{\omega l} \phi_{\omega l}(r) P_{l}(\cos\theta).
\end{eqnarray}

The absorption cross section depends on the flux of particles that enter the black hole through the effective potential. Hence,  we can introduce the four-current density vector as follows
\begin{eqnarray}\label{Q35}
J^{\mu}=\frac{i}{2}(\Phi^{*}\Delta^{\mu}\Phi-\Phi\Delta^{\mu}\Phi^{*}),
\end{eqnarray}
and the above equation satisfies the conservation law, that is
\begin{eqnarray}\label{Q36}
\Delta_{\alpha}J^{\alpha}=0.
\end{eqnarray}

By substituting Eq.(\ref{Q34}) into Eq.(\ref{Q35}) under the boundary condition Eq.(\ref{E_nled3}), we obtain the four-current density vector by surface integral as
\begin{eqnarray}\label{Q37}
N(r)=-\int_{\Sigma}r^{2}J^{r}d\Omega =-\frac{\pi}{\omega}\sum_{l=0}^{\infty} (2l+1)(1-\left|e^{2 i \delta_{l}
}|^2\right),
\end{eqnarray}
where $N(r)$  is the flux that passes the surface $\Sigma$ with a constant radius $r$ and $d\Omega=\sin\theta d\theta d\varphi$. The flux is a constant, and when we consider the stationary scenarios, $N$ (minus) represents the particles passing through the potential and entering the black hole \cite{Unruh1976}. Besides, we have used the orthogonality of Legendre polynomials, i.e.,
\begin{eqnarray}\label{Q38}
\int_{-1}^{1} P_{l}(x) P_{l'}(x) dx =\frac{2}{(2l+1)}\delta_{l l'}.
\end{eqnarray}
where $x=\cos\theta$.	
Furthermore, the absorption cross section $\sigma_{abs}$ is defined as the ratio of the particle flux $\left|N\right|$ to the plane wave incident current $\omega$. Hence, the absorption cross section can be written as
\begin{eqnarray}\label{Q39}
\begin{aligned}
& \sigma_{abs}(\omega)\equiv \frac{\left|N\right|}{\omega}= \frac{\pi}{\omega^{2}}\sum_{l=0}^{\infty}(2l+1)(1-\left|e^{2 i \delta_{l}
}|^2\right) \\
& ~~~~~~~~~~=\frac{\pi}{\omega^{2}} \sum_{l=0}^{\infty} (2l+1)\left|T_{\omega l}\right|^{2},
\end{aligned}
\end{eqnarray}
and the partial absorption cross section can be expressed as
\begin{eqnarray}\label{Q40}
\sigma_{l}(\omega)= \frac{\pi}{\omega^{2}}(2l+1)(1-\left|e^{2 i \delta_{l}}|^2\right)= \frac{\pi}{\omega^{2}}(2l+1) \left|T_{\omega l}\right|^{2}.
\end{eqnarray}

In order to study the effects of Rastall and string parameters on the absorption cross section of the scalar field, we need to calculate the phase shift $\delta_{l}$, that is, the transmission coefficient. In this paper, we use the WKB approximation to obtain the transmission coefficient $T_{\omega}$. Assuming that the probability of the incident plane wave is equal to 1, Eq.(\ref{E_nled4}) can be expressed as
\begin{eqnarray}\label{Q41}
\left|R_{\omega l}\right|^2+\left|T_{\omega l}\right|^2=1.
\end{eqnarray}

The transmission probability of different multipole numbers $l$ can be obtained with the help of the sixth-order WKB method,
\begin{eqnarray}\label{Q42}
1- \left|R_{\omega l}\right|^2=\left|T_{\omega l}\right|^2,
\end{eqnarray}
with
\begin{eqnarray}\label{Q43}
R_{\omega l}=(1+e^{2i \pi \alpha})^{-\frac{1}{2}},
\end{eqnarray}
where $\alpha$ is obtained by
\begin{eqnarray}\label{Q44}
\alpha-i \frac{(\omega^{2}-V_{0})}{\sqrt{-2V_{0}^{''}}} -\sum_{i=2}^{i=6} \Lambda_{i}(K)=0.
\end{eqnarray}

In Eq.(\ref{Q44}), $V_{0}$ represents the maximum value of the potential at $r=r_{0}$, and the prime denotes the derivative of the potential at $r=r_{0}$ with respect to $r^{*}$. Moreover, $\Lambda_{i}(K)$ indicates a higher-order correction of the WKB method, which depends on $K$ and the  2$i$ order derivative of the potential at its maximum position \cite{Iyer1987,Konoplya2003}.

Specifically, we express the third-order method as follows,
\begin{eqnarray}\label{Q45}
\begin{aligned}
&\Lambda_{2}=\frac{1}{\sqrt{-2V^{(2)}_{0}}}[\frac{1}{8}(\frac{V^{(4)}_{0}}{V^{(2)}_{0}}(b^2+\frac{1}{4})-\frac{1}{288}(\frac{V^{(3)}_{0}}{V^{(2)}_{0}})^2(7+60b^2))]\\
&\Lambda_{3}=\frac{n+\frac{1}{2}}{-2V^{(2)}_{0}}[\frac{5}{6912}(\frac{V^{(3)}_{0}}{V^{(2)}_{0}})^4(77+188b^2) \\
& -\frac{1}{384}(\frac{(V^{(3)}_{0})^2V^{(4)}_{0}}{(V^{(2)}_{0})^3})(51+100b^2)+\frac{1}{2304}(\frac{V^{(4)}_{0}}{V^{(2)}_{0}})^2(67+ \\
& 68b^2)-\frac{1}{288}(\frac{V^{(6)}_{0}}{V^{(2)}_{0}})(5+4b^2)+\frac{1}{288}(\frac{V^{(3)}_{0}V^{(5)}_{0}}{(V^{(2)}_{0})^2})(19+28b^2)].
\end{aligned}
\end{eqnarray}

In Eqs. (\ref{Q45}), the superscripts (2,3,4,5,6) of the effective potential represent the corresponding differentials with respect to the tortoise coordinate $r_{*}$, and $b=n+\frac{1}{2}$. Besides, since the specific expressions of $\Lambda_{4}(K)$, $\Lambda_{5}(K)$ and $\Lambda_{6}(K)$ are overly cumbersome (see Ref. \cite {Konoplya2003}), they will not be described in detail here. In addition, during the calculation, we find that when the Rastall parameter $\beta$ is set as a fraction, the results and figures of the WKB approximation calculation are more accurate than when $\beta$ is set as a decimal \cite{Cai2020}. This phenomenon can be attributed to the term $r^{\frac{4\beta}{2\beta-1}}$ in the metric $f(r)$. Hence, in order to maintain the consistency of the data, we choose the fractional form of $\beta$ throughout the paper.

From Fig. \ref{FIG3}, we can compare the effects of the Rastall parameter and the string parameter on the partial absorption cross section when the Rastall parameter is positive. The results are shown in the left plot, where different values of the string parameter $a$ are chosen, the corresponding partial absorption cross section first starts at zero, then reaches to a maximum value, and finally decreases to almost the same value with increasing $\omega$. Furthermore, it is easy to see that as the string parameter $a$ increases, the partial absorption increases and its peak position shifts to the left. When we fix the string parameter and change the Rastall parameter, one can get  that the peak value of the partial absorption cross section increases as the Ratall parameter $\beta$ increases.

\begin{figure*}\label{FIG3}
	\centering
	  \includegraphics[height=5cm]{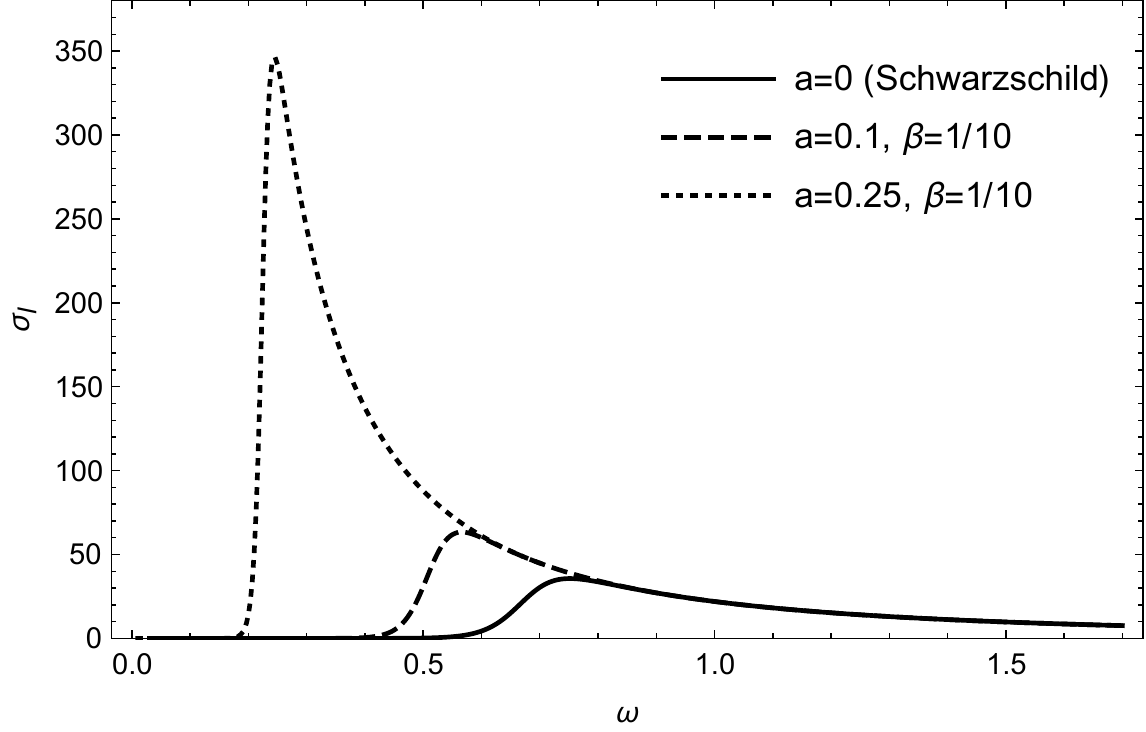}\quad
	  \includegraphics[height=5cm]{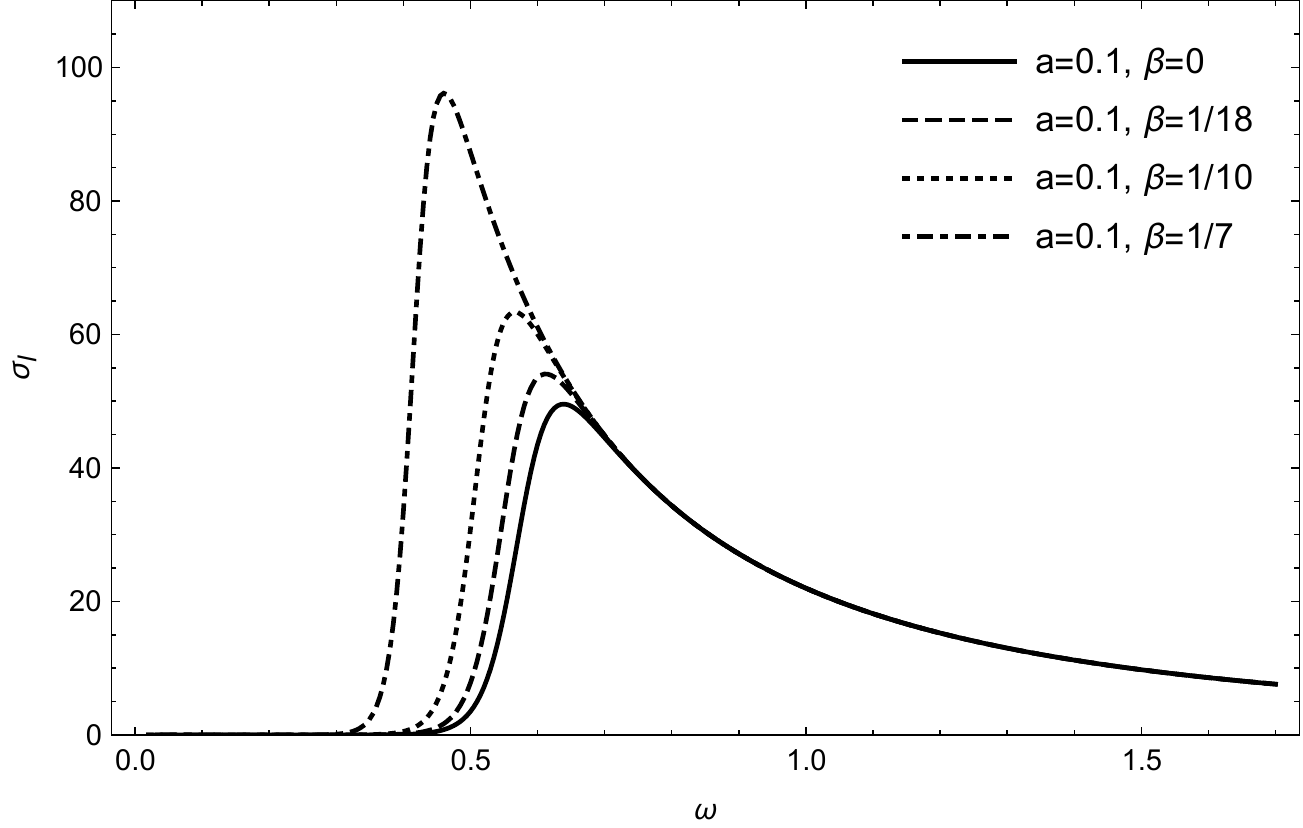}\quad
	  \includegraphics[height=5cm]{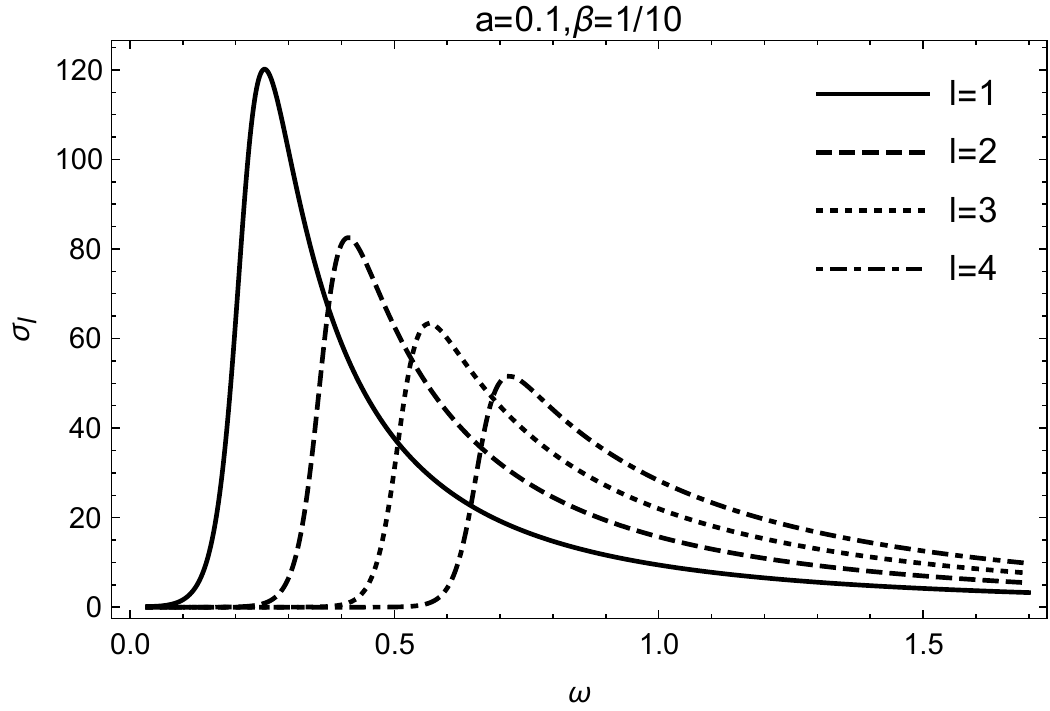}\quad
	\caption{\label{FIG3} For $\beta>0$ and $M=1$, the partial absorption cross section for the variable $a$ at $\beta =\frac{1}{10}$, $l=3$ in the left panel and the  variable  $\beta$ at $a=0.1$, $l=3$ in the right panel. The lower plot is that the partial absorption cross section changes with the multipole number $l$ at $a=0.1,\beta=\frac{1}{10}$.
}\end{figure*}

\begin{figure*}\label{FIG4}
	\centering
	  \includegraphics[height=5.1cm]{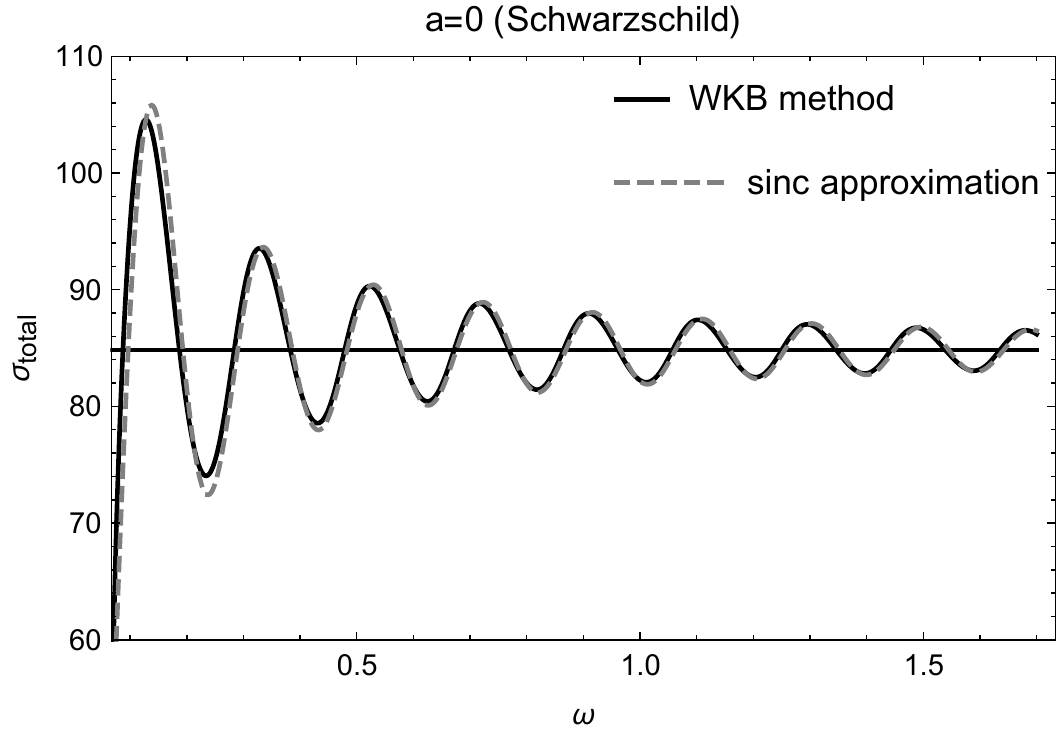}\quad
	  \includegraphics[height=5cm]{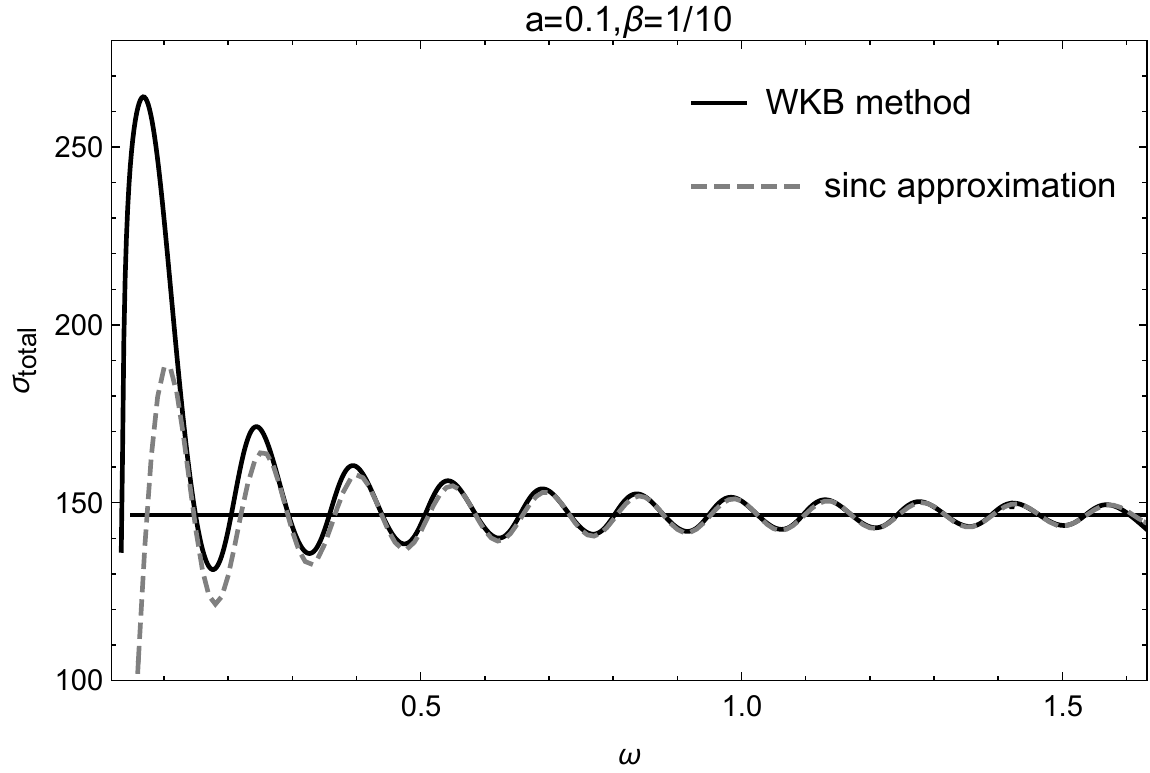}\quad
	  \includegraphics[height=5cm]{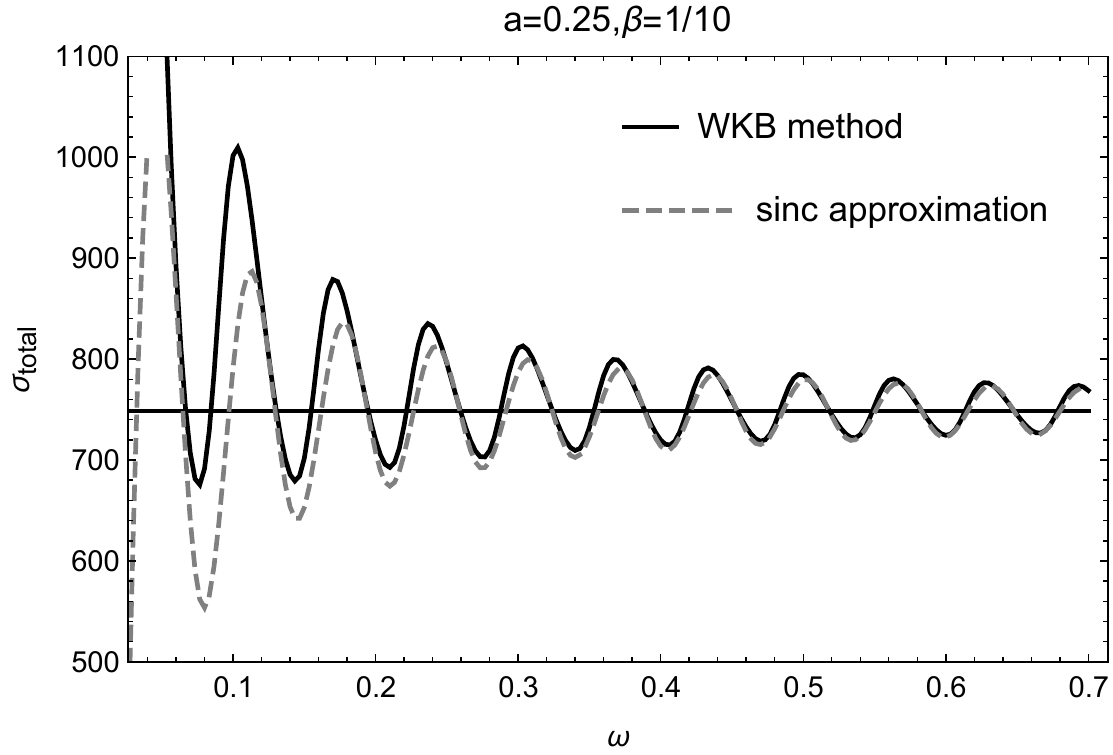}\quad
	\caption{\label{FIG4}For $\beta>0$ and $M=1$, the total absorption cross section for the variable $a$ at $\beta= \frac{1}{10}$, $a=0$ recovers to the Schwarzschild black hole as the contrast.
}\end{figure*}

\begin{figure*}\label{FIG5}
	\centering
	\includegraphics[height=5.1cm]{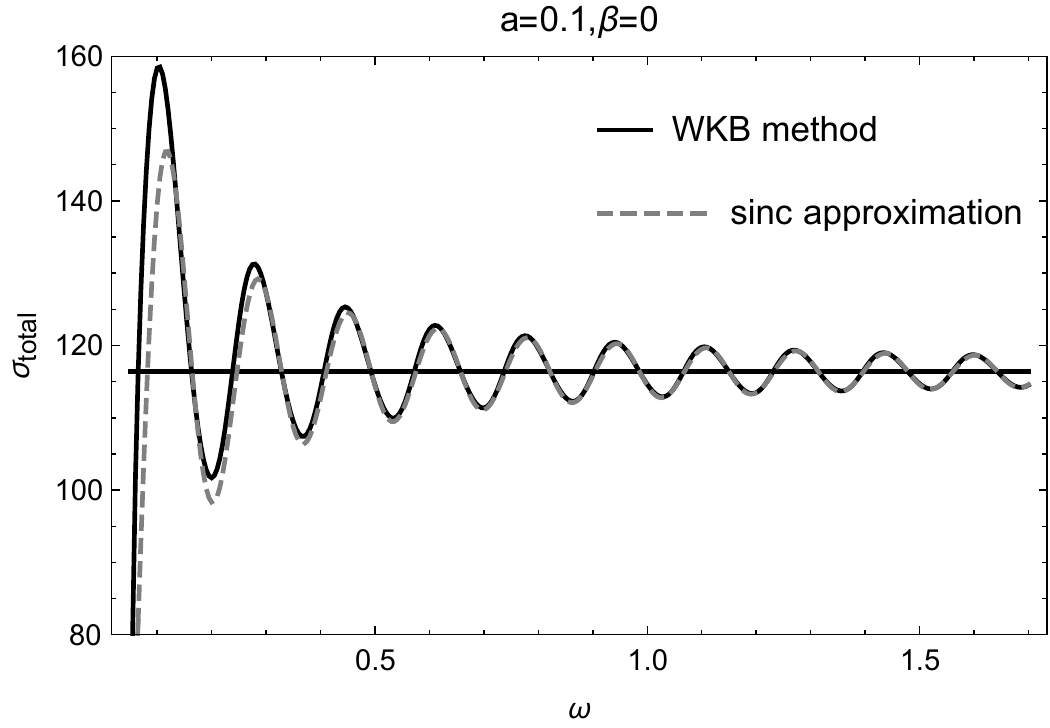}\quad
	\includegraphics[height=5cm]{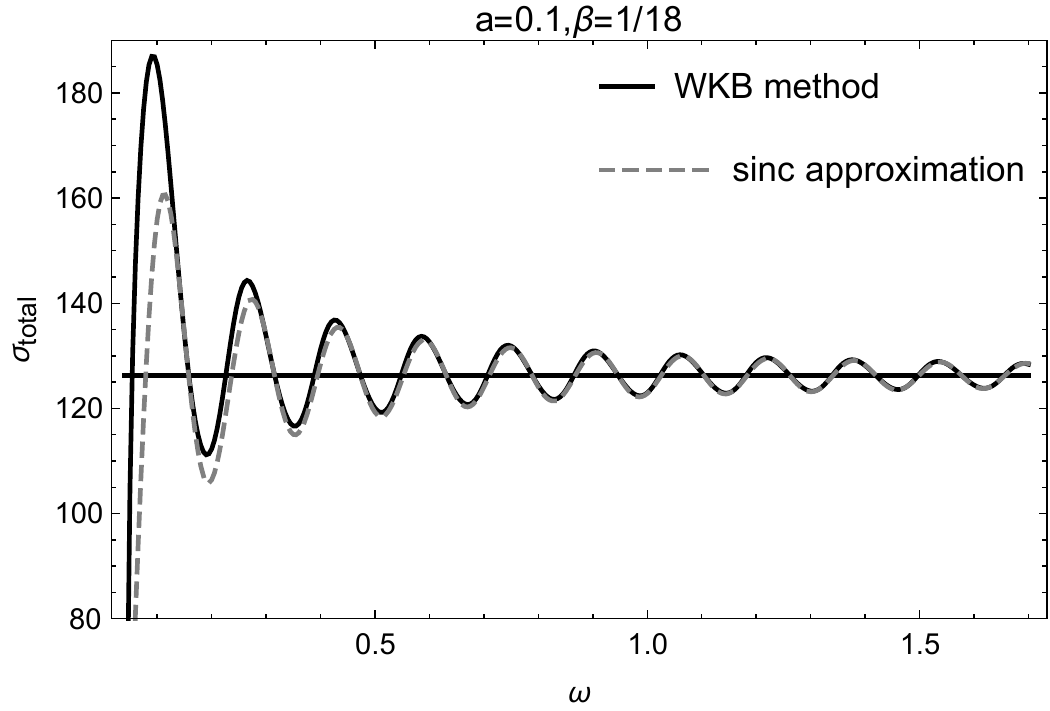}\quad
	\includegraphics[height=5cm]{F9.pdf}\quad
	\includegraphics[height=5cm]{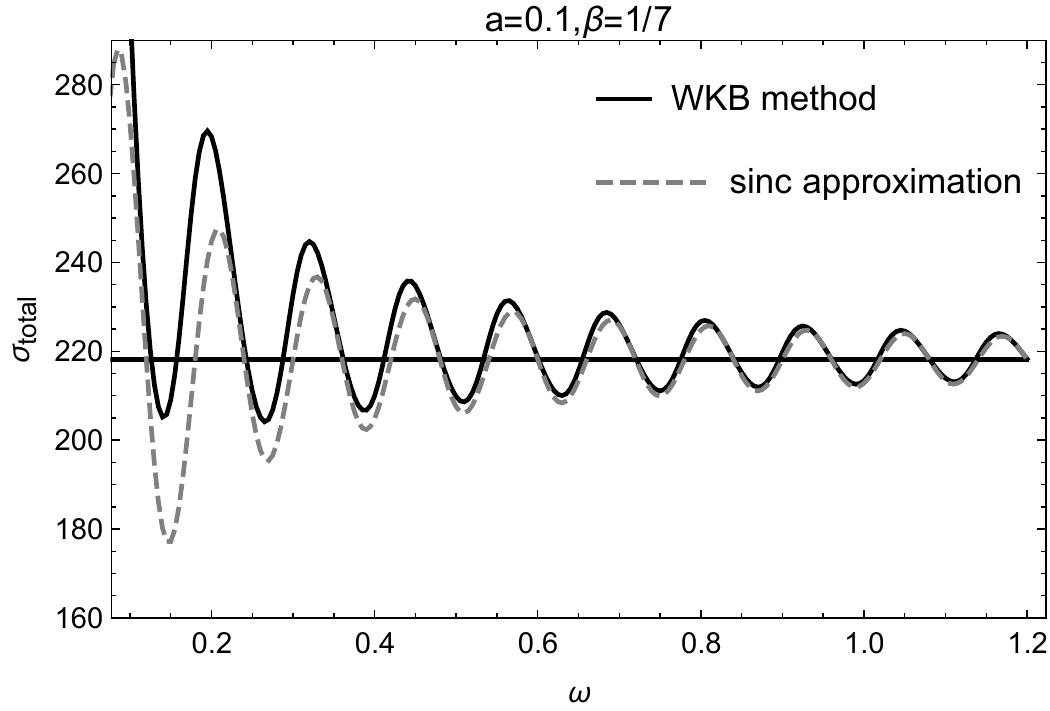}\quad
	\caption{\label{FIG5}For $\beta>0$ and $M=1$ , the total absorption cross section for the variable $\beta$ at $a =0.1$,  and  $\beta =0$ recovers to the Einstein gravity.
}\end{figure*}

\begin{figure*}\label{FIG6}
	\centering
	\includegraphics[height=5cm]{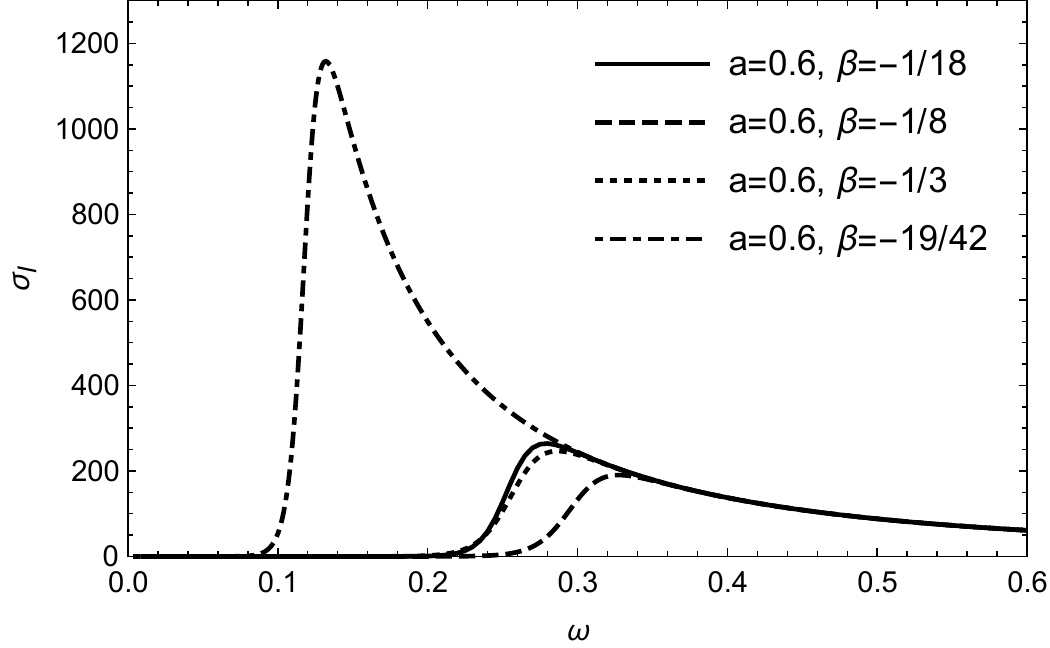}\quad
	\includegraphics[height=5.1cm]{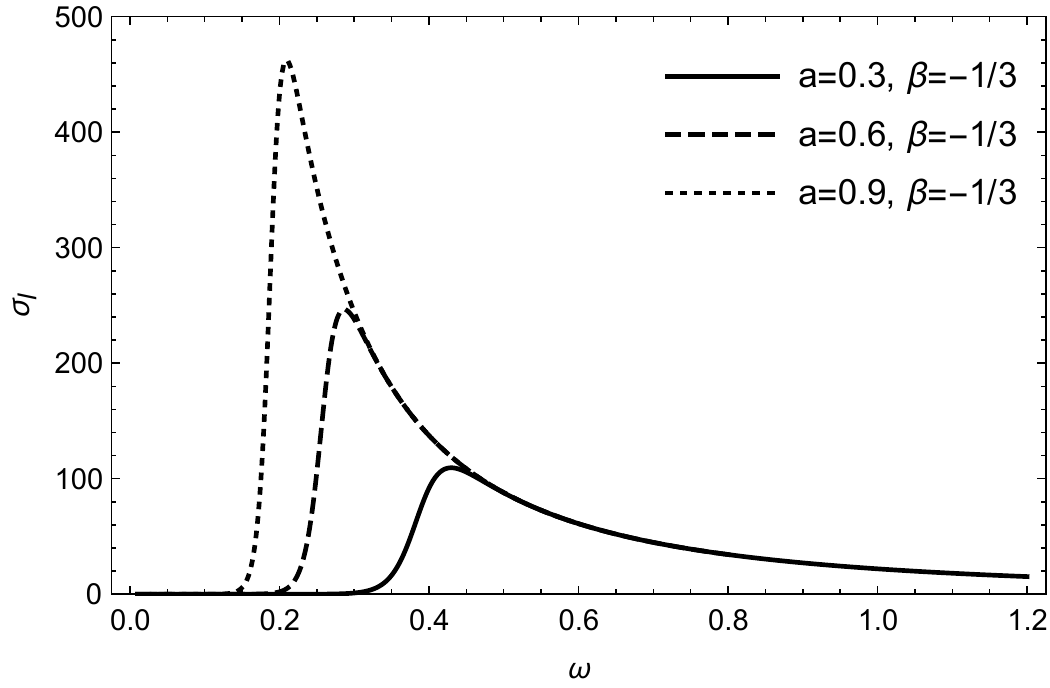}\quad
	\includegraphics[height=5cm]{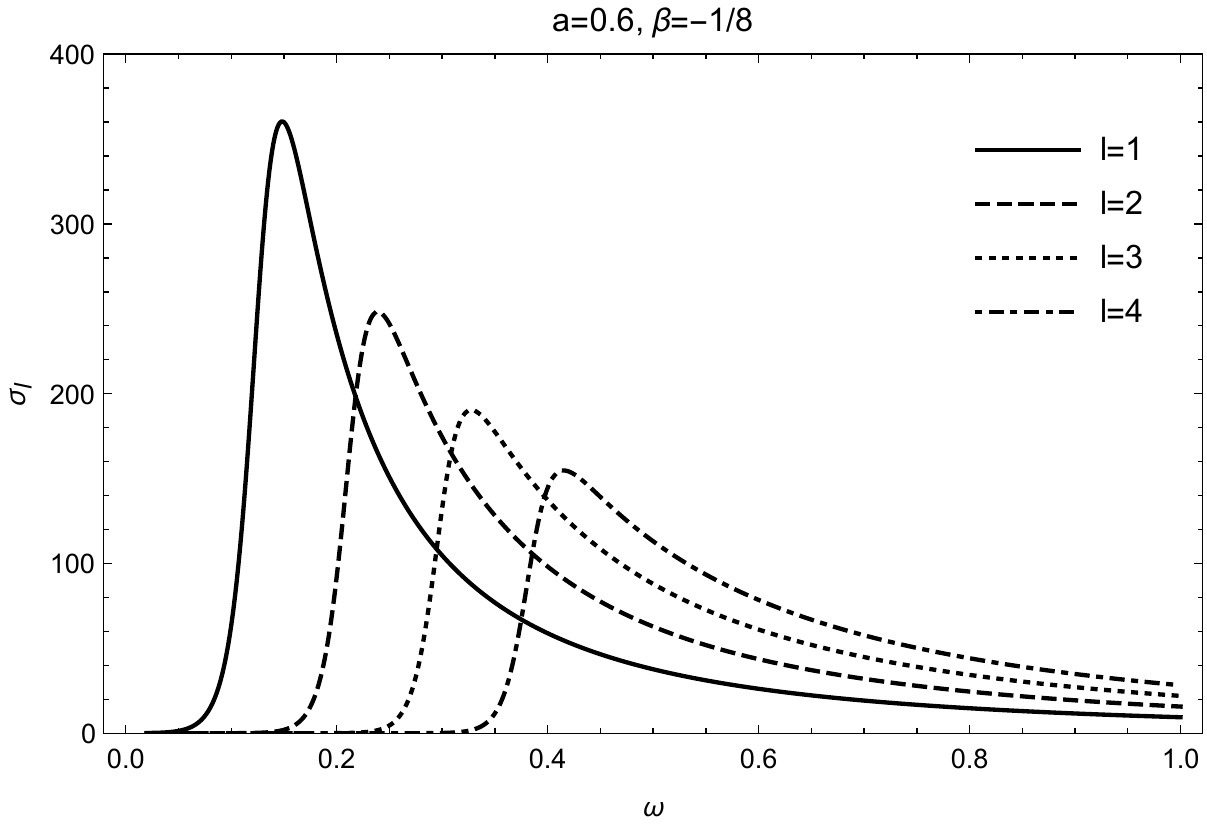}\quad
	\caption{\label{FIG6}For $\beta>0$ and $M=1$, the partial absorption cross section for the variable $a$ at $\beta =\frac{1}{10}$, $l=3$ in the left panel and the  variable is $\beta$ at $a=0.1$, $l=3$  in the right panel. The lower plot is that the partial absorption cross section change with the multipole number $l$ at $a=0.1,\beta=\frac{1}{10}$.
}\end{figure*}

In Fig. \ref{FIG4} we present the total absorption cross section of a Schwarzschild black hole surrounded by a cloud of strings in Rastall gravity for different values of the string parameter, where $l$ goes from $0$ to $10$ and $\beta=\frac{1}{10}$. Specifically, the horizontal solid line represents the geometric capture section. As shown in Fig. 4, the dashed curve is the sinc approximation result, and the solid curve is the partial wave result using the sixth-order WKB approximation. We show that increasing the parameter $a$ results in incrementing the absorption cross section. We also notice that the two curves are significantly different at small values of frequency. Moreover, as the string parameter increases, the difference is more pronounced and the range of oscillation amplitudes is significantly wider. However, in the high frequency regime, the total absorption cross sections obtained by these two methods are in good agreement and converge to the geometric capture cross section. In Fig. \ref{FIG5}, our results show that when we fix the value of $a$ and increase $\beta$, the absorption cross section increases. The difference between the two curves also increases significantly in the low frequency regime due to the Rastall parameter.

As shown in Fig. \ref{FIG6}, we describe the behavior of the partial absorption cross section for different parameters obtained by the sixth-order WKB method when $\beta < 0$. From the left figure, where the Rastall parameter is treated as a variable and the parameter $a$ is fixed, we observe that the partial absorption cross section does not monotonically increase as the Rastall parameter $\beta $ decreases. The partial cross sections intersect in the range of $0.2<\omega< 0.3$ due to the effective potential.  Therefore, the variation trend of the partial cross section is $\text{\text{'N'}}$ type. We also observe in the right plot that when we increase the string parameter, the partial cross section increases monotonically. Moreover, the peak position of the partial cross section is evidently shifted to the left.
Finally, we observe that as the multipole number $l$ increases, the partial cross section decreases and its peak position shifts to the right.

\begin{figure*}\label{FIG7}
	\centering
	  \includegraphics[height=5cm]{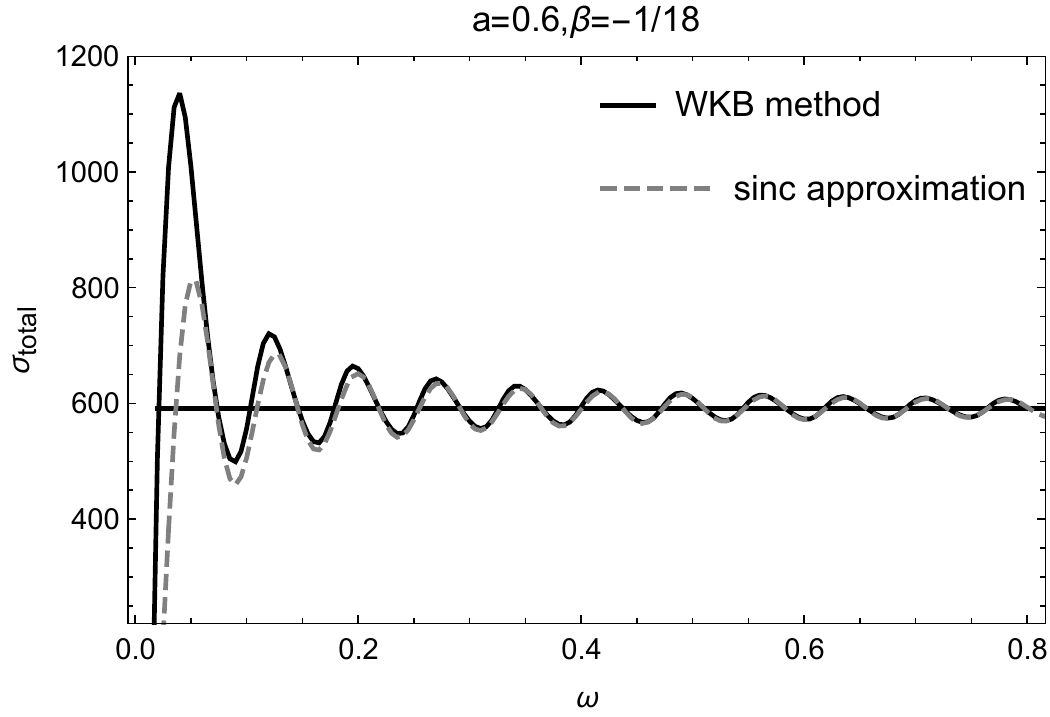} \quad
	  \includegraphics[height=5cm]{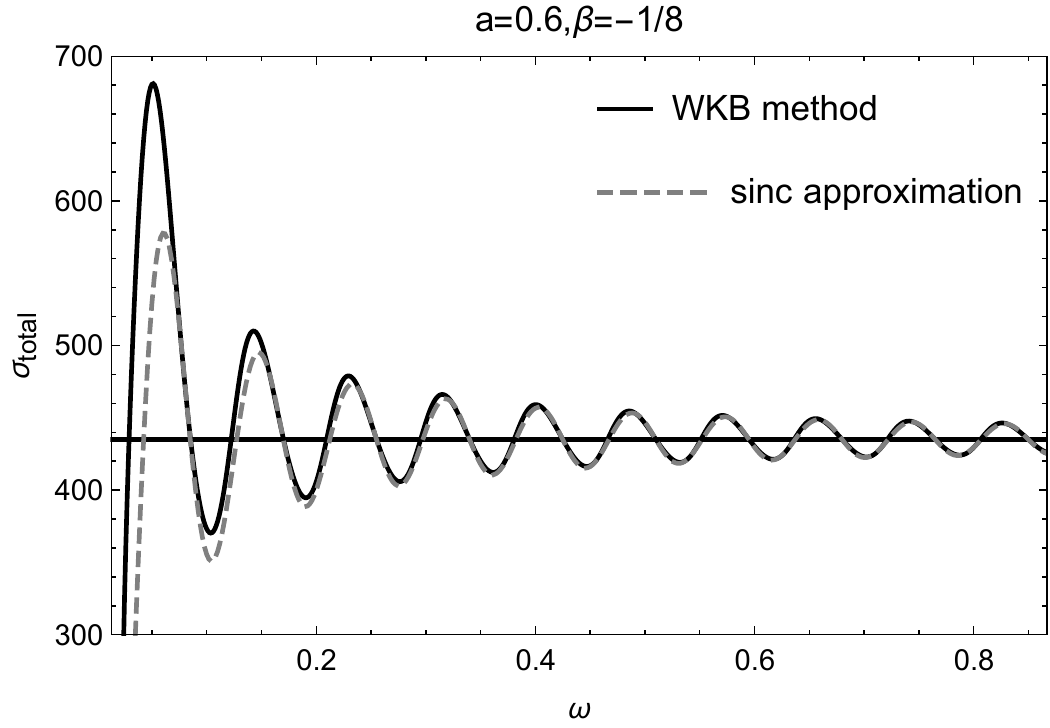}\quad
	  \includegraphics[height=5cm]{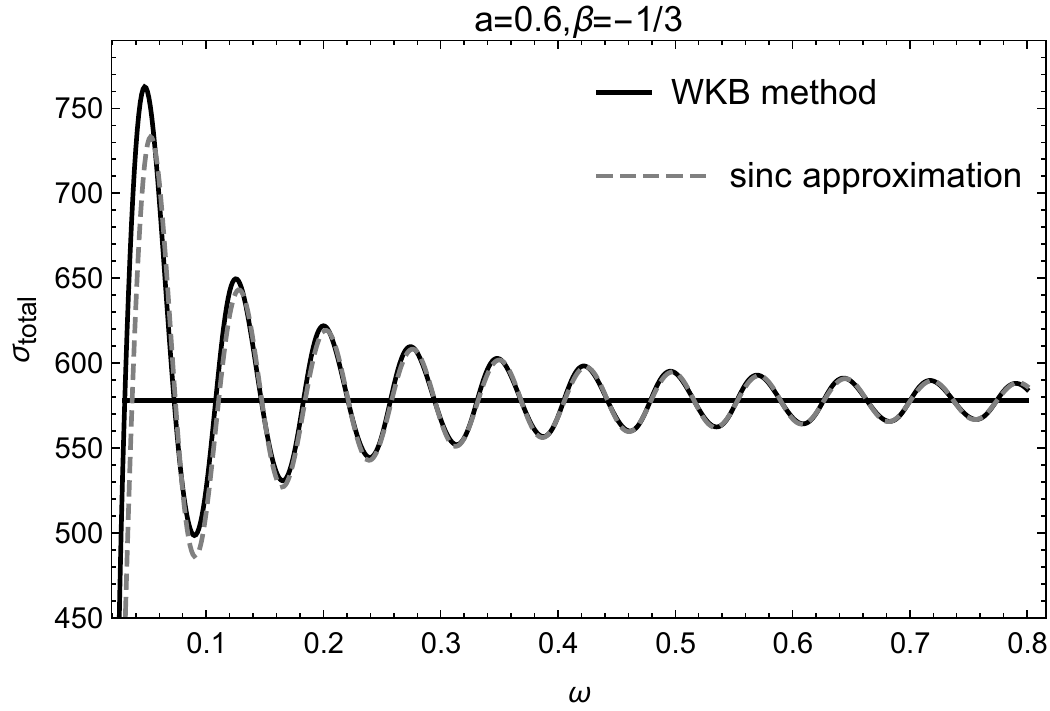}\quad
	  \includegraphics[height=5cm]{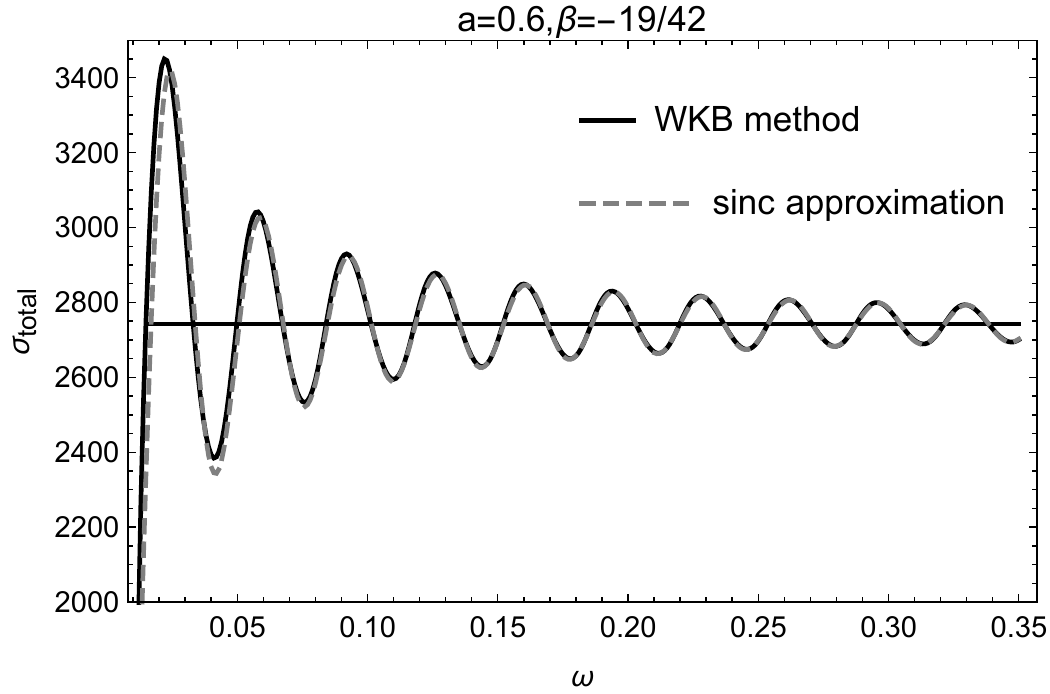}\quad
	\caption{\label{FIG7}For $\beta<0$  and $M=1$, the total absorption cross section for the variable $\beta$  with fixed string parameter $a=0.6$.
}\end{figure*}

In Fig. \ref{FIG7} we give the total absorption cross sections of the massless scalar field by varying the Rastall parameter $\beta$ ($\beta<0$) and fixing the string parameter $a = 0.6$. We can observe that the change of the total absorption cross section as a function of $\beta$ is similar to that of the partial absorption cross section. This is because the higher the potential barrier, the more particles are scattered back to the black hole by the potential barrier. In addition, we can see that when we reduce the Rastall parameter to $-0.5$, the difference between the solid curve and the dashed curve gradually decreases.

\begin{figure*}\label{FIG8}
	\centering
	  \includegraphics[height=5cm]{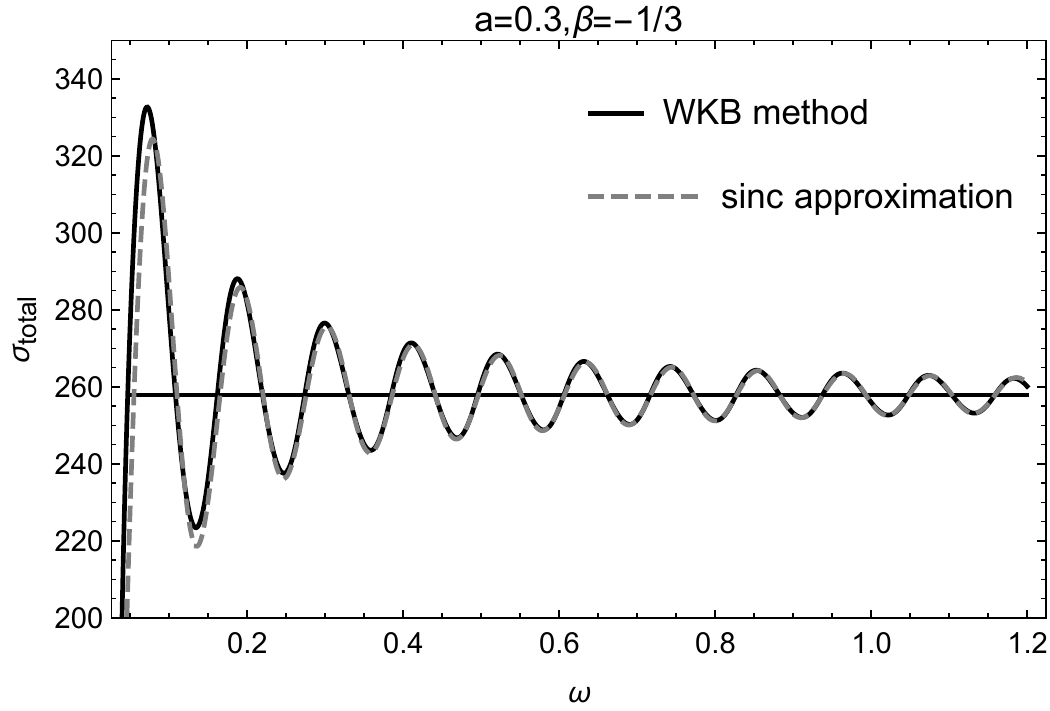}\quad
	  \includegraphics[height=5cm]{F20.pdf}\quad
	  \includegraphics[height=5cm]{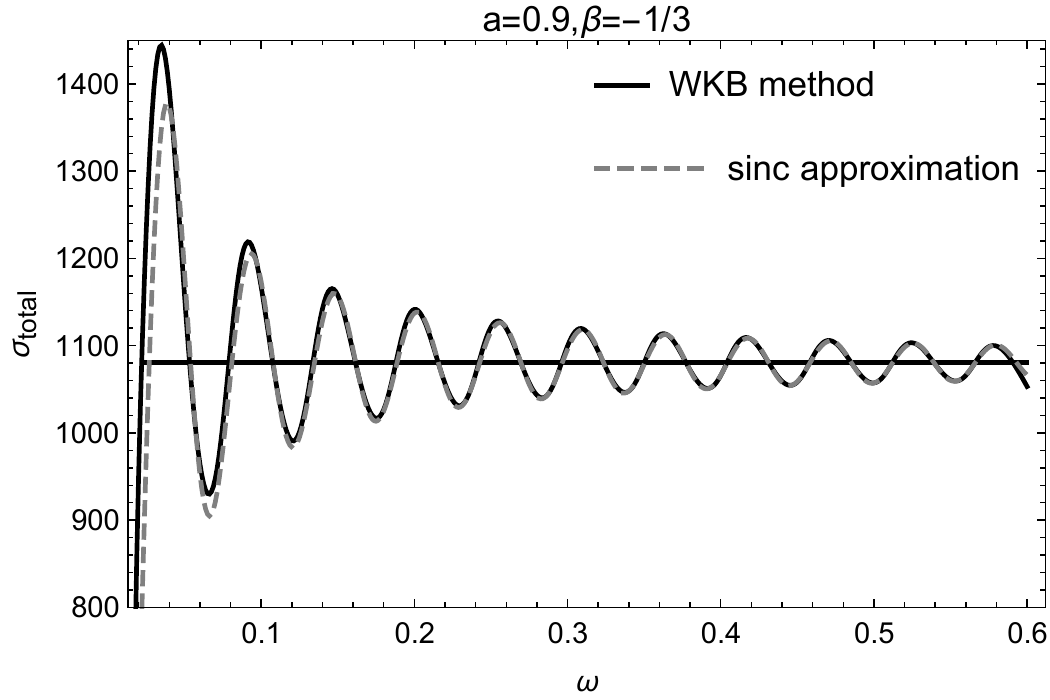}\quad
	\caption{\label{FIG8}For $\beta<0$ and $M=1$, the total absorption cross section for the variable $a$  with fixed  string parameter $\beta=-\frac{1}{3}$.
}\end{figure*}

In Fig. \ref{FIG8} we present the total absorption cross section of the massless scalar field when $\beta<0$, changing the string parameters and fixing $\beta=-\frac{1}{3}$. It can be observed that the difference between the two curves is the smallest at the low-frequency limit compared to the above three cases. Furthermore, when $\omega$ is large, the total absorption cross section as a function of $\omega$ goes into the capture cross section. We also notice that the total absorption cross section, as well as the oscillation amplitude, increases with increasing string parameters.

\section{Hawking radiation }

In this section, we employ the sixth-order WKB method to calculate the Hawking radiation for massless scalar fields. Furthermore, we analyze the effects of the string and the Rastall parameters on Hawking radiation in the background of a Schwarzschild black hole surrounded by a cloud of strings in Rastall gravity.

A black hole behaves almost in the same way as a black body, emitting particles when its temperature is proportional to the surface gravity \cite{Hawking1975}. Hawking also presented that black holes can radiate particles in the form of thermal. This is due to the quantum tunneling effect created by the vacuum fluctuations near the event horizon of the black hole. Therefore, if we consider quantum effects and the laws of thermodynamics are satisfied, black holes can produce radiation. This phenomenon is known as Hawking radiation.

The Hawking radiation calculated by the gray-body factor has the following expression \cite{Sharif:2020cus,Sharif:2021sow}
\begin{eqnarray}\label{Q46}
\frac{dE}{dt}=\sum_{l}N_{l}|T_{\omega l}|^{2}\frac{\omega}{\text{exp}(\omega/T_{BH})- 1}\frac{d\omega}{2\pi},
\end{eqnarray}
where $N_{l}$ is the multiplicity that depends only on the black hole dimension. Moreover, for the massless scalar field in a four-dimensional black hole, $l$ and $N_{l}$ satisfy the condition $N_{l}=2l+1$. $T_{\omega l}$ denotes the above gray-body factor and $T_{BH}$ represents the Hawking temperature.
Specifically, the Hawking temperature of static spherically symmetric spacetime can be written as
\begin{eqnarray}\label{Q47}
T_{BH}=\frac{1}{4\pi}f'(r)\bigg |_{r=r_{h}}.
\end{eqnarray}

By substituting Eq.(\ref{Q6}) into Eq.(\ref{Q47}), we can obtain
\begin{eqnarray}\label{Q48}
T_{BH}=\frac{1}{4\pi r_{h}}\bigg (1+\frac{a(1-2\beta)r^{\frac{4\beta}{1-2\beta}}_{h}}{4\beta-1}\bigg),
\end{eqnarray}
where $f(r_{h})=0$ and $r_{h}$ is the radius of the event horizon. Besides, the string  and Rastall parameters need to satisfy the previous parameter range, i.e., $-0.5< \beta< \frac{1}{6}$ and $0 \leq a < 1$. By substituting Eq.(\ref{Q48}) and  $N_{l}=2l+1$ into Eq.(\ref{Q46}), we can further obtain the Hawking power emission spectrum
\begin{eqnarray}\label{Q49}
\frac{d^{2}E}{dt d\omega }=\frac{1}{2\pi}\sum_{l}\frac{(2l+1)|T_{\omega l}|^{2} \omega}{e^{\omega/T_{BH}}-1}.
\end{eqnarray}

Fig.\ref {FIG9} compares the effects of parameters $a$ and $\beta$ on the Hawking power emission spectrum of the massless scalar wave when $\beta$ is non-negative. We can clearly observe in the upper panel that for a given $l$ and $\beta$, increasing the parameter $a$ depresses the power emission spectrum. Moreover, the peak power emission spectrum gradually shifts to low frequencies as $a$ increases. It is clear from the middle panel that when we fix  $l$ and $a$, but increase the parameter $\beta$, the peak power emission spectrum gradually decreases and moves to low frequencies. As the multipole number $l$ increases, we can get from the lower panel that for a massless scalar field, the power emission spectrum decreases and the peak position shifts towards high frequencies. In conclusion, the parameters $a$, $\beta$ and $l$ suppress the power emission spectrum. Besides, it is easy to see that if the values of parameters   $a$ and $b$ are chosen larger, the lifespan of the black hole will be longer.

This trait is more easily observed in Fig.\ref{FIG10}, which plots the effects of parameters $a$, $\beta$ and $l$ on the power emission rate (as a function of $\omega$) for the scalar wave in the range $\beta<0$. From the upper figure we can see that when we increase the parameter $a$, the power emission spectrum decreases. That is, under the condition that $\beta$ is constant, the increase of the string parameter $a$ leads to a decrease in the energy emission rate, thus making the lifetime of the black hole longer.
Furthermore, we also observe in the center panel that with decreasing Rastall parameter, for fixed $l$ and $a$, the peak value of the power emission rate increases and then decreases, and the peak position first shifts to high frequency and then moves to low frequency.  Finally, we fix the two parameters $a=0.6$ and $\beta=-\frac{1}{3}$ and analyze the effects of the multipole number $l$ in the lower panel. It is clear that a larger multipole number results in a lower power emission spectrum. Besides, it is worth noting that the low multipole number $l$ dominates the energy emission rate, while the contribution of the high multipole number $l$ is extremely small and thus negligible.

\begin{figure*}\label{FIG9}
	\centering
	  \includegraphics[height=5cm]{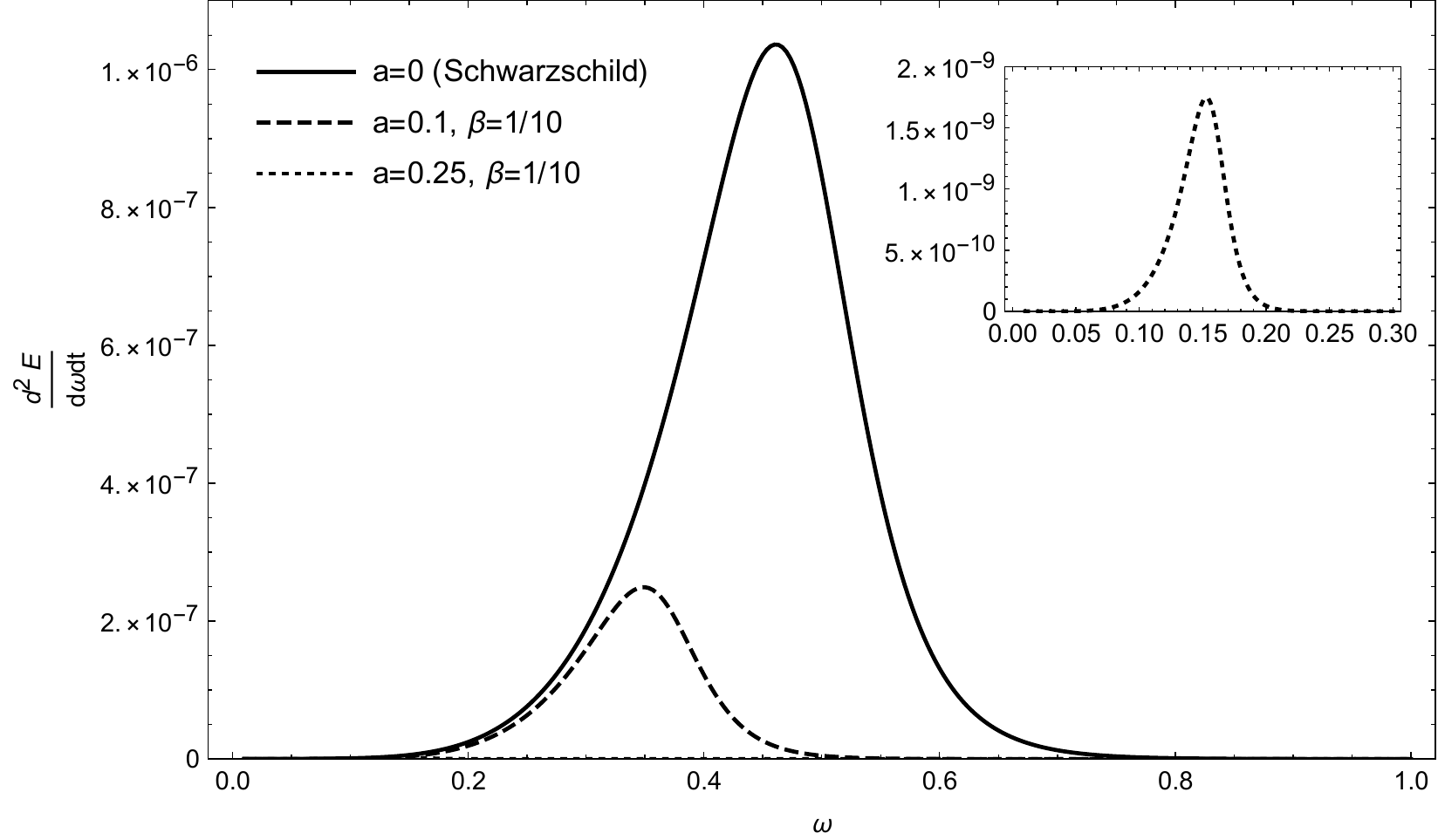}\quad
	  \includegraphics[height=5.18cm]{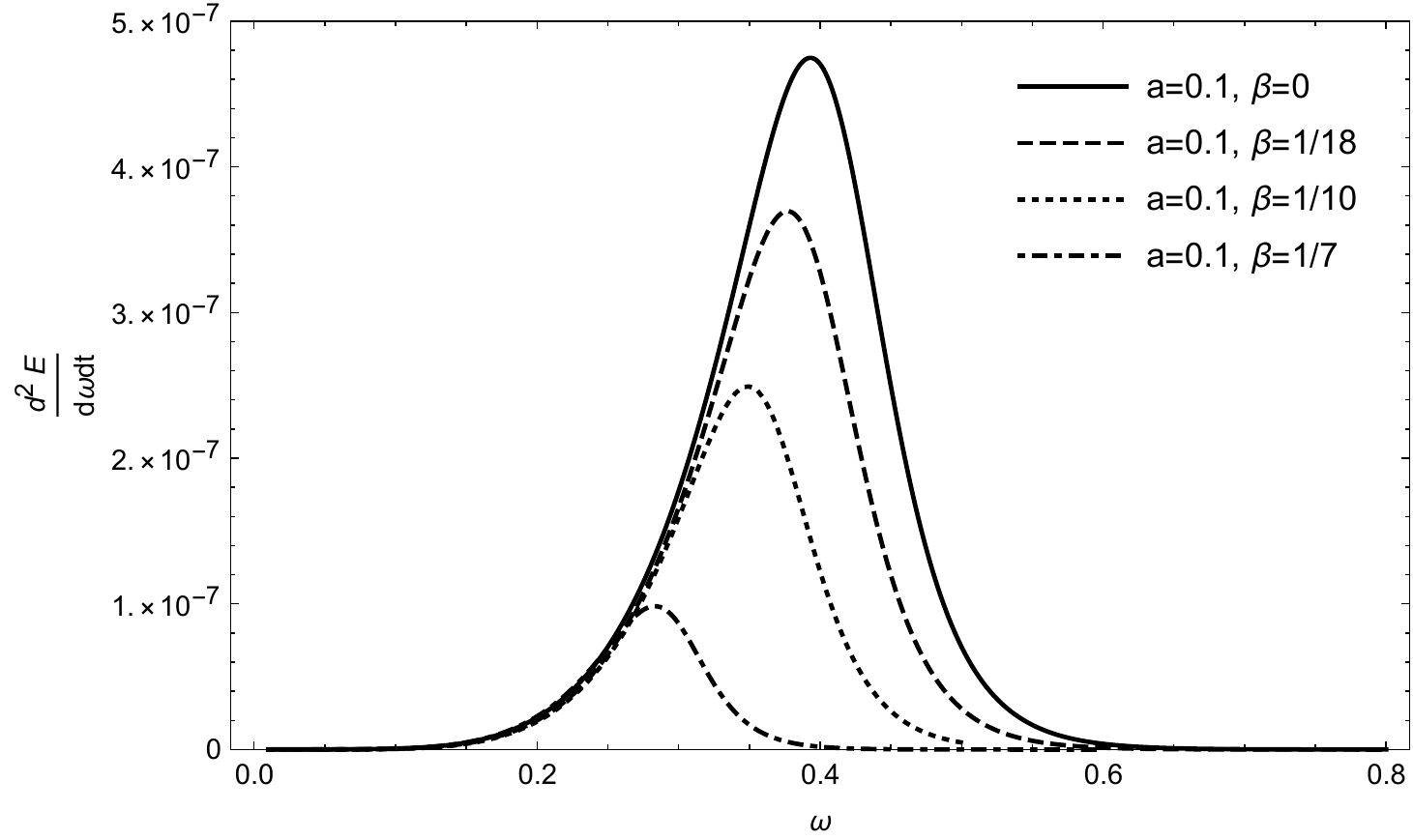}\quad
	  \includegraphics[height=5cm]{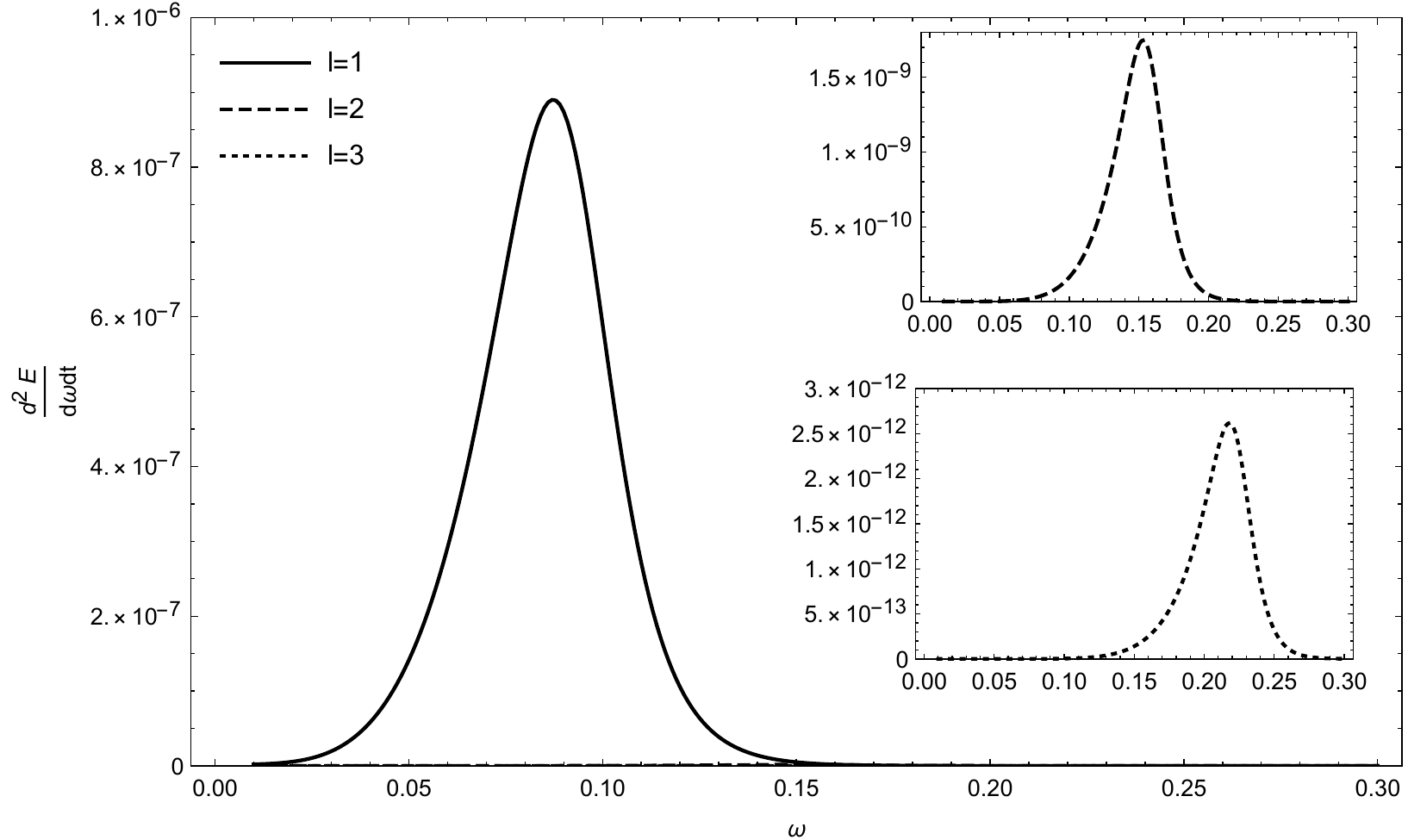}\quad
	\caption{\label{FIG9}For $\beta>0$  and $M=1$, the upper-figure is the power emission spectra taken $a$ as the variable at $\beta=\frac{1}{10}$, $l=2$. The center-figure is the power emission spectra taken $\beta$ as the variable at $a=0.1$, $l=2$. The lower-figure is the power emission spectra taken $l$ as the variable at $a=0.25$ and $\beta=\frac{1}{10}$.
}\end{figure*}

\begin{figure*}\label{FIG10}
	\centering
	  \includegraphics[height=5cm]{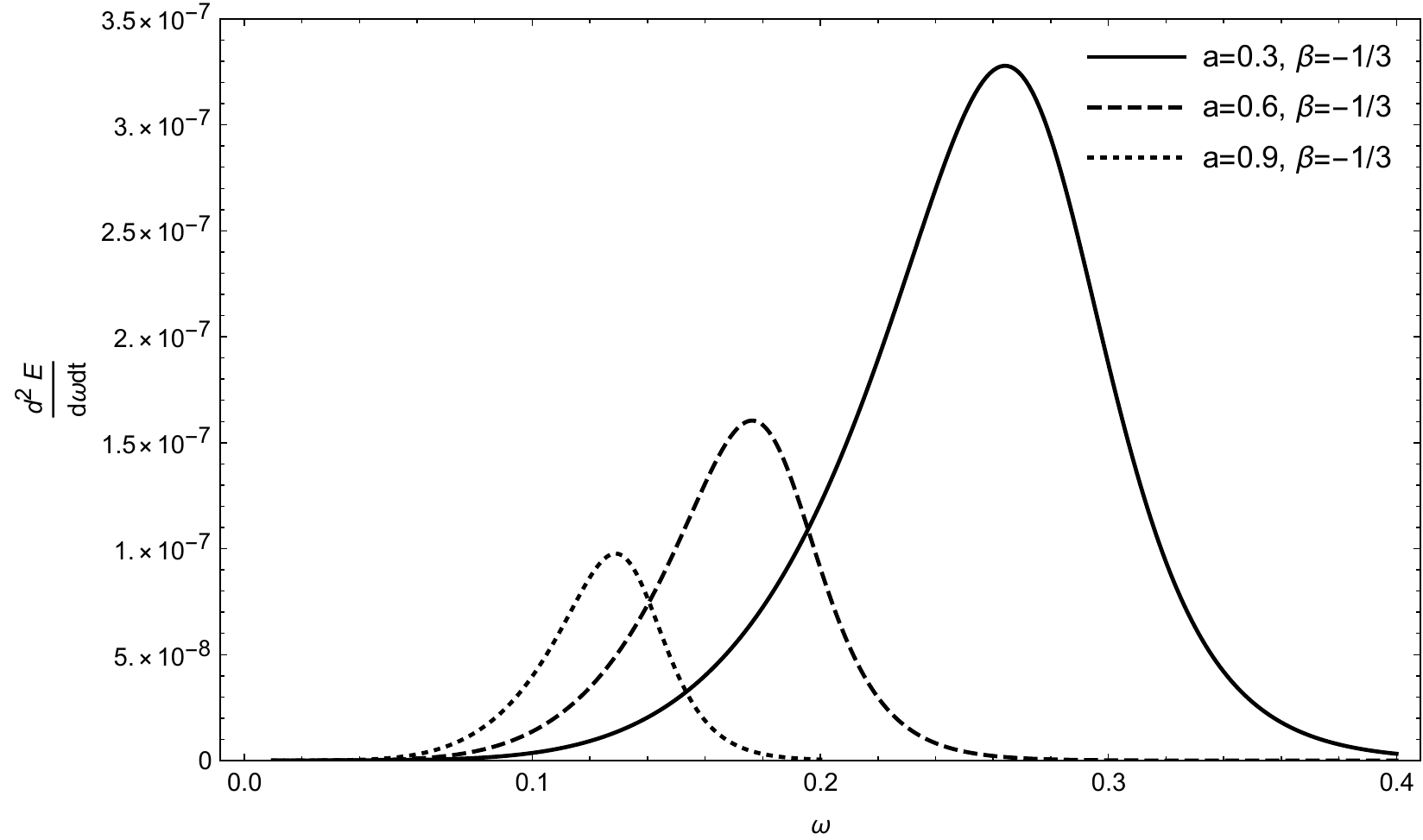}\quad
	  \includegraphics[height=4.90cm]{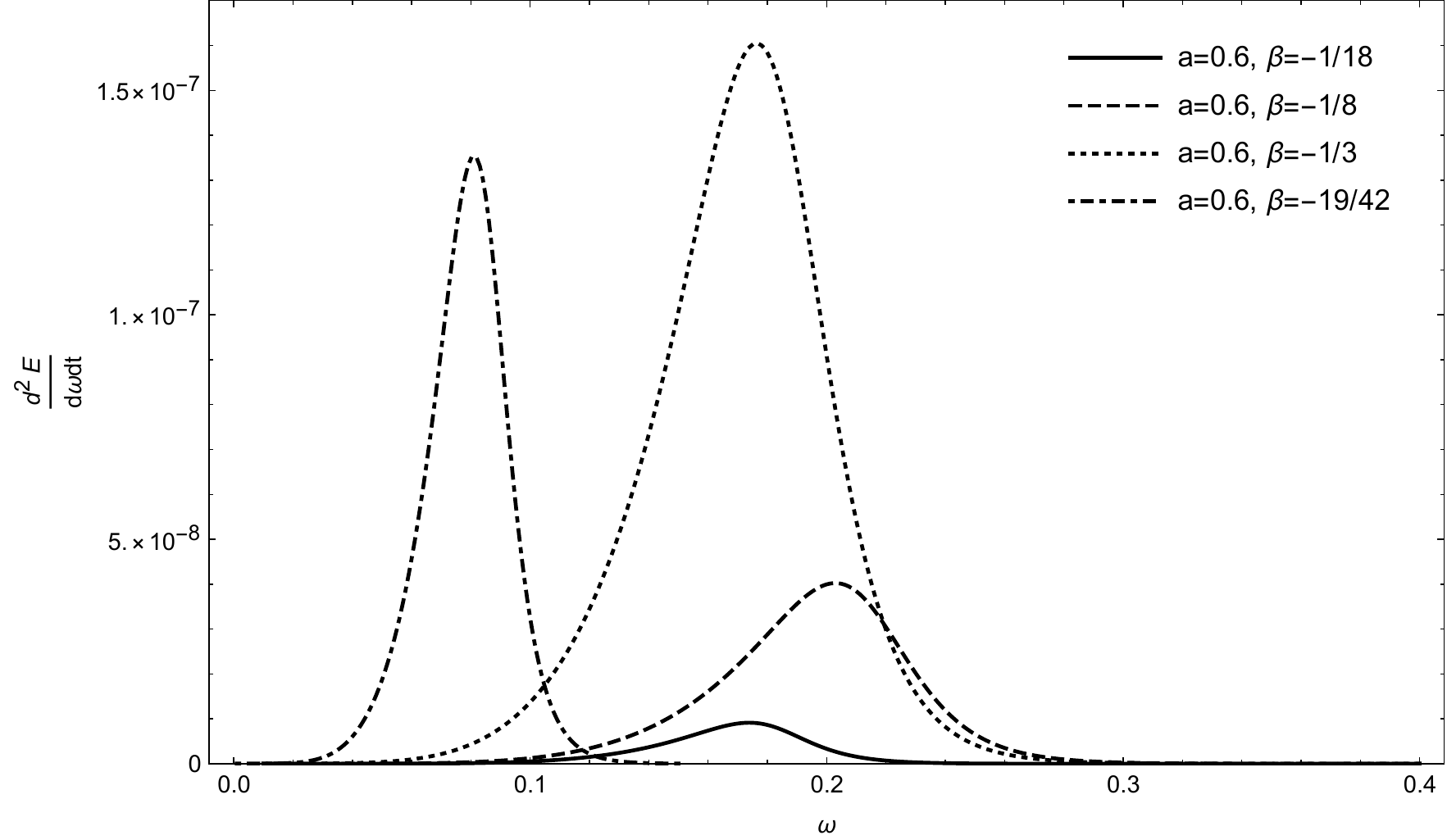}\quad
	  \includegraphics[height=5cm]{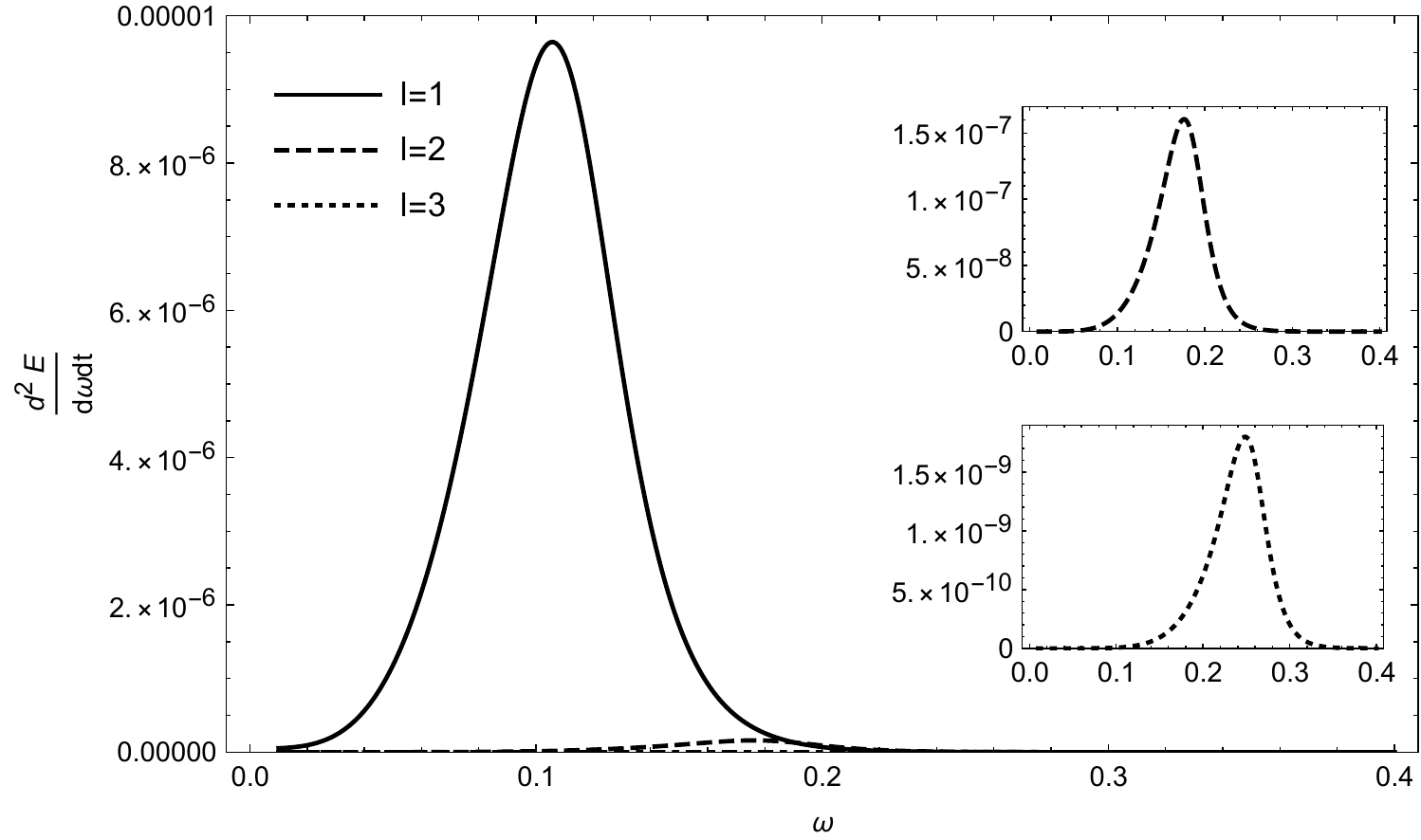}\quad
	\caption{\label{FIG10}For $\beta<0$  and $M=1$, the upper-figure is the power emission spectra taken $a$ as the variable at $\beta=-\frac{1}{3}$, $l=2$. The center-figure is the power emission spectra taken $\beta$ as the variable at $a=0.6$, $l=2$. The lower-figure is the power emission spectra taken $l$ as the variable at $a=0.6$ and $\beta=-\frac{1}{3}$.
}\end{figure*}

\section{Conclusion and discussion}
\label{sect:conclusion}

In the previous sections, we have comprehensively studied the black hole shadow, absorption cross section and power emission spectrum of Hawking radiation for the massless scalar field in a Schwarzschild black hole surrounded by a cloud of strings in Rastall gravity. The ranges of the string parameter and Rastall parameter are chosen according to the effective potential in the context of the scalar field. Notably, we have calculated the absorption cross section and Hawking radiation with the help of the sixth-order WKB method.

First, in Figs.\ref{FIG1} and \ref{FIG2}, we  carefully analyzed   the effective potential for different values of parameters $a$, $\beta$ and $l$.
For $\beta>0$, the parameters $a$ and $\beta$ depress the barrier of the effective potential, and the waves do not reflect.  For $\beta<0$, $a$ reduces the barrier height when $\beta$ is fixed, whereas the effective potentials intersect when $\beta$ varies. Moreover, we studied the shadow  and  photon sphere radii caused by the curved light ray. Because we consider the black hole to be static spherically symmetric, the radii of the photon sphere and shadow are constant. In other words, the black hole shadow has spherical symmetry. Besides, the radius of the photon sphere increases as the parameter $a$ increases. However, when we consider $\beta$ as a variable, the photon sphere and shadow radii fluctuate abnormally. The reason is that when the Rastall parameter is less than zero, the metric $f(r)$ changes abnormally.

Second, with the help of the sixth-order WKB method, we calculated the absorption cross section of the scalar field in detail. To compare the accuracy of the sixth-order WKB, we also presented the results of the sinc approximation with the geometric capture cross section as a reference. From Figs.\ref{FIG3}, \ref{FIG4} and \ref{FIG5}, we can clearly observe that larger values of the parameters $a$ and $\beta$ enhance the partial or total absorption cross section when $\beta>0$. However, in the low frequency range, when $a$ or $\beta$ is set to a larger value, the results calculated by the two methods are quite different. Furthermore, in Figs.\ref{FIG6},  \ref{FIG7} and \ref{FIG8}, we plotted the partial and total absorption cross sections when $\beta$<0. Unlike the case where $\beta$ is positive, the absorption cross section does not always grow as the Rastall parameter decreases. Since the potential barrier reflects waves, the change in the absorption cross section is exactly the opposite of the change in the potential barrier. Hence, as $\beta$ decreases, the total absorption cross section first increases, then decreases and finally increases again. It is worth mentioning that the smaller the value of $\beta$, the smaller the difference between the two approximations. Very importantly, in the mid-high frequency region, the total absorption cross section and the sinc approximation are in good agreement and in all cases oscillate around the geometric capture cross section $\sigma_{geo}$.

Finally, we investigated the energy emission rate of Hawking radiation. Specifically, the power emission rate is affected by the string parameter, the Rastall parameter as well as the multipole number. In Fig.\ref{FIG9}, we found that both $a$ and $\beta$ suppress the power emission spectrum, and the peak position shifts to a lower energy region. Moreover, the multipole number $l$ also significantly depresses the power emission spectra whereas the peak position shifts to the higher frequency regime. The case of $\beta<0$ is also similar to the case of $\beta>0$ above, except the case where $\beta$ varies and $a$ is fixed. As the Rastall parameter decreases, the power emission spectrum first increases and then decreases, at the same time, the peak position first moves to the higher frequency  region and then enters the lower energy region.

\subsection*{acknowledgments}
This work was supported partly by the National Natural Science Foundation of China (Grants No. 12065012, No. 12065013), Yunnan High-level Talent Training Support Plan Young \& Elite Talents Project (Grants No. YNWR-QNBJ-2018-360) and the Fund for Reserve Talents of Young and Middle-aged Academic and Technical Leaders of Yunnan Province (Grant No. 2018HB006).	

\subsection*{Data Availability Statement}
This manuscript has no associated data	or the data will not be deposited. [Authors’ comment: This present study	is a theoretical work.]

\end{document}